\documentclass[aps,pra,superscriptaddress,letterpaper,longbibliography,notitlepage]{revtex4-1}
\usepackage{graphicx}
\usepackage{amsmath}
\usepackage{esint}
\usepackage{verbatim}
\usepackage{color}
\usepackage{SIunits}
\usepackage{hyperref}

\usepackage{color}

\usepackage{multirow}

\definecolor{light-gray}{gray}{0.80}
\usepackage{soul}
\usepackage{braket}

\newcommand{\gsq}{{g}^{\prime}}
\newcommand{\gsqzpf}{\tilde{g}^{\prime}}
\newcommand{\g}{{g}}

\newcommand{\xzpfk}{x_{\text{zpf,$k$}}}
\newcommand{\xzpf}{x_{\text{zpf}}}
\newcommand{\xkeq}{x_{\text{$k$,eq}}}
\newcommand{\xk}{x_{k}}
\newcommand{\meff}{m_{\text{eff}}}
\newcommand{\omegam}{\omega_{\text{m}}}
\newcommand{\omegamk}{\omega_{\text{m,$k$}}}
\newcommand{\vecb}[1]{\mathbf{#1}}
\newcommand{\Htot}{{\hat{\mathcal{H}}_{\text{tot}}}}
\newcommand{\Hint}{\hat{\mathcal{H}}_{\text{int}}}
\newcommand{\Hopt}{\hat{\mathcal{H}}_{\text{opt}}}
\newcommand{\Hmec}{\hat{\mathcal{H}}_{\text{mec}}}
\newcommand{\Hnot}{\hat{\mathcal{H}}_{\text{0}}}

\sethlcolor{yellow}

\begin{document}

\title{Design of a quasi-2D photonic crystal optomechanical cavity with tunable, large $x^2$-coupling}

\author{Mahmoud Kalaee}
\affiliation{Kavli Nanoscience Institute and Thomas J. Watson, Sr., Laboratory of Applied Physics, California Institute of Technology, Pasadena, CA 91125, USA}
\affiliation{Institute for Quantum Information and Matter, California Institute of Technology, Pasadena, CA 91125, USA}
\author{Taofiq K. Para\"iso}
\affiliation{Max Planck Institute for the Science of Light, G\"unther-Scharowsky-Stra\ss e 1/Bau 24, D-91058 Erlangen, Germany}
\affiliation{Kavli Nanoscience Institute and Thomas J. Watson, Sr., Laboratory of Applied Physics, California Institute of Technology, Pasadena, California 91125, USA}
\author{Hannes Pfeifer}
\affiliation{Max Planck Institute for the Science of Light, G\"unther-Scharowsky-Stra\ss e 1/Bau 24, D-91058 Erlangen, Germany}
\author{Oskar Painter}
\email{opainter@caltech.edu}
\affiliation{Kavli Nanoscience Institute and Thomas J. Watson, Sr., Laboratory of Applied Physics, California Institute of Technology, Pasadena, CA 91125, USA}
\affiliation{Institute for Quantum Information and Matter, California Institute of Technology, Pasadena, CA 91125, USA}  

\date{\today}


\begin{abstract}
We present the optical and mechanical design of a mechanically compliant quasi-two-dimensional photonic crystal cavity formed from thin-film silicon in which a pair of linear nanoscale slots are used to create two coupled high-$Q$ optical resonances.  The optical cavity supermodes, whose frequencies are designed to lie in the $1500$~nm wavelength band, are shown to interact strongly with mechanical resonances of the structure whose frequencies range from a few MHz to a few GHz.  Depending upon the symmetry of the mechanical modes and the symmetry of the slot sizes, we show that the optomechanical coupling between the optical supermodes can be either linear or quadratic in the mechanical displacement amplitude.  Tuning of the nanoscale slot size is also shown to adjust the magnitude and sign of the cavity supermode splitting $2J$, enabling near-resonant motional scattering between the two optical supermodes and greatly enhancing the $x^2$-coupling strength.  Specifically, for the fundamental flexural mode of the central nanobeam of the structure at $10$~MHz the per-phonon linear cross-mode coupling rate is calculated to be $\tilde{g}_{+-}/2\pi = 1$~MHz, corresponding to a per-phonon $x^2$-coupling rate of $\tilde{g}'/2\pi=1$~kHz for a mode splitting $2J/2\pi = 1$~GHz which is greater than the radiation-limited supermode linewidths.
%These multimoded electro-opto-mechanical resonators constitute a new class of devices with high potentialities for both the measurement of quantum signatures in mesoscopic devices and the development of integrated optical information processing devices.
\end{abstract}

\maketitle

%\ocis{(130.0130) Integrated optics; (230.5298) Photonic crystals; (230.4555) Coupled resonators; (230.4685) Optical microelectromechanical devices.} 

%%%%%%%%%%%%%%%%%%%%%%% References %%%%%%%%%%%%%%%%%%%%%%%%%

\section{Introduction}
\label{sec:intro}
Multimode optomechanical systems consisting of three or more optical and mechanical modes have recently received growing interest within the field of cavity optomechanics~\cite{RevModPhys.86.1391}. One of the main original motivations for multimoded systems was the prospect of realizing a quantum non-demolition (QND) measurements of mechanical energy or phonon number using position-squared optomechanical coupling arising from the novel dispersion of a multimode optical system~\cite{jayich,thompson,ludwig2012}. While QND measurements turn out to be very challenging in such a set-up~\cite{ludwig2012,miao}, multimode optomechanical systems are promising for studying a variety of other interesting phenomena, such as three-mode parametric instability~\cite{braginsky2001parametric,braginsky2002analysis,braginsky2002low}, multimode optomechanically induced transparency \cite{Ojanen2014,Fan-Tang-cascaded-eit}, synchronization of mechanical oscillators \cite{heinrich2011collective,PhysRevLett.109.233906,PhysRevLett.111.213902,PhysRevLett.112.014101} and the generation of sub-Poissonian statistics~\cite{Xu2013,Flayac2015,Lemonde2014,Lorch2015}.  Multimode optomechanical systems may also find application in displacement sensing~\cite{kippenberg2005analysis,krause-accelerometer} and optical information processing~\cite{schmidt2012optomechanical}, where they can be used as optical filters~\cite{Winger:chip-scale}, switches or delay lines \cite{Stannigel_Lukin,chang2011slowing,Safavi-Naeini2011-EIT,Weis2010-OMIT}.  Given the strong interest and opportunity in multimode cavity optomechanics, a number of new experimental platforms have been developed.  These include membrane-in-the-middle setups~\cite{thompson,sankey2010strong,Lee2015,Chen2015}, nanofabricated chip-scale resonators~\cite{grudinin2010phonon,Fan-Tang-cascaded-eit,Doolin2014,Chen2015}, hybrid multimode microwave circuits~\cite{Massel2012}, and gravitational wave detectors~\cite{Evans2015}.   

\begin{figure}[t]
\centering
\includegraphics[width=0.75\columnwidth]{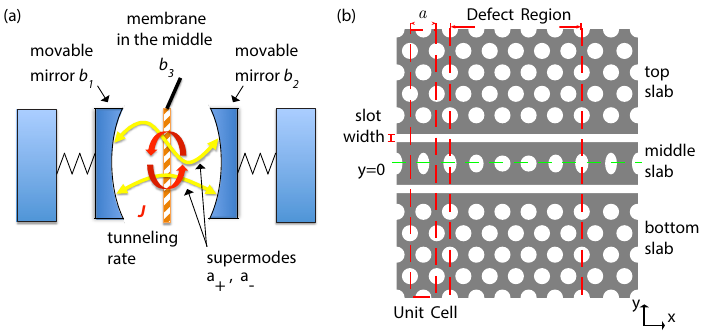}
\caption{(a) Schematic of a multimode membrane-in-the-middle optomechanical system consisting of a central movable membrane ($b_3$) and two movable end mirrors ($b_1$, $b_2$).  Due to tunneling (rate $J$) of light through the partially transmitting central membrane, the left and right individual optical cavity modes ($a_1$, $a_2$; not shown) hybridize into the supermodes $a_+$ and $a_-$. (b) Schematic of a photonic crystal implementation of a similar multimode optomechanical system.  The structure consists of a pair of top and bottom photonic crystal slabs which are separated from a central photonic crystal slab by nanoscale air slots.  A pair of optical waveguide modes localize around each nanoscale slot, propagating along the axial direction ($x$) of the structure.  By varying the photonic crystal unit cell along the length of the structure, one can form optical cavity modes which are localized to a central ``defect'' region of the structure.  Light in the cavity modes surrounding the top and bottom nanoscale air slots ($a_1$, $a_2$) tunnel across the central photonic crystal slab, forming hybridized supermodes ($a_+$, $a_-$).  Mechanical motion of the structure includes in-plane flexural motion of the central nanobeam ($b_3$), the top photonic crystal slab ($b_1$), and the bottom photonic crystal slab ($b_2$).}
\label{fig:design}
\end{figure}

In a recent experimental work, we realized a silicon photonic crystal optomechanical cavity capable of very large position-squared optomechanical coupling~\cite{Paraiso2015}.  Here, we present details of the cavity design and explore the range of possible optomechanical interactions in such a device.  As shown schematically in Fig.~\ref{fig:design} the structure consists of a double-waveguide photonic crystal cavity in which two individual waveguide cavity modes are coupled by photon tunneling through a mechanically compliant element.  The double-waveguide cavity localizes two optical resonances at telecommunication wavelengths $\lambda=$1550~nm with high quality factors $Q>10^5$. The optical modes couple efficiently to mechanical modes with frequencies $\omegam/2\pi$ ranging from $6$~MHz up to $1$~GHz. We propose a tuning scheme based on electrostatic actuation~\cite{Winger:chip-scale}, which allows for control of the optical tunneling rate between the two slotted waveguide modes as well as the supermode frequencies. This provides a direct dynamical control over the linear and quadratic optomechanical coupling strengths.

The paper is organized as follows. We start in Sec.~\ref{sec:theory} with a brief introduction of the theory of multimode optomechanics. We consider a generic Hamiltonian describing a system of two optical modes and a series of mechanical modes. Coupling between the optical modes occurs via mechanically independent photon tunneling, and via absorption or emission of mechanical phonons. By considering specific symmetries of the optical and mechanical modes we show how different orders of coupling in mechancial amplitude can occur.  In Sec.~\ref{sec:design} we detail the design of the specific double-slotted photonic crystal cavity of interest to our work.  We show that by controlling the widths of the waveguide slots of the structure it is possible to adjust the photon tunneling rate of the optical modes, and to even completely suppress this tunneling. In Sec.~\ref{sec:mechanicaldesign} we analyze the mechanical resonances of the photonic crystal cavity and identify several mechanical modes with frequencies ranging from MHz to GHz that have the appropriate symmetry for realizing large optomechanical coupling.  Finally, in Sec.~\ref{sec:coupling} we employ a perturbation theory to calculate the strength of the linear self-mode, linear cross-mode, and position-squared coupling a number of the different mechanical resonances of the photonic crystal structure.  

\section{Multimode Hamiltonian}
\label{sec:theory}
We consider a multimode cavity optomechanical system consisting of two spatially separated optical modes $a_{1,2}$ of frequencies $\omega_{1,2}$ and independent mechanical modes $b_k$ of frequencies $\omega_{m,k}$. The individual optical cavity modes are coupled to each other through photon tunneling at a rate $J$. Conceptually, our structure can be viewed as an on-chip generalization of the membrane-in-middle setups~\cite{thompson,ludwig2012} in a chip-scale architecture, as it is schematically represented in Fig.~\ref{fig:design}. The corresponding Hamiltonian can be written as,

\begin{align}
\begin{split}
\Htot &= \Hopt + \Hmec + \Hint \, ,\\
\Hopt &=\hbar \omega_{1} \hat{a}_{1}^{\dagger} \hat{a}_{1} + \hbar \omega_{2} \hat{a}_{2}^{\dagger} \hat{a}_{2} + \hbar J (\hat{a}_{1}^{\dagger} \hat{a}_{2} + \hat{a}_{2}^{\dagger} \hat{a}_{1}), \\
\Hmec &= \hbar \sum_{k} \omegamk \hat{b}_k^{\dagger} \hat{b}_k\, , \\
\Hint &= \hbar \sum_{i,j,k} \g_{ij,k} \hat{a}_{i}^{\dagger} \hat{a}_{j} (\hat{b}_k +\hat{b}_k^{\dagger} )\xzpfk \, ,
\end{split}
\label{eq:geninteraction}
\end{align}

\noindent where $\g_{ij,k}$ are the coupling strengths between the optical modes $a_i$, $a_j $ and the mechanical mode $b_k$. In the particular case of symmetric optical cavities, $a_1$ and $a_2$ are degenerate with $\omega_1=\omega_2=\omega_0$. Since the individual optical modes are spatially separated, we can to a good approximation neglect the terms proportional to $g_{ij,k}$ with $i \neq j$ and therefore write $g_{ij,k}$ simply as $g_{i,k}$.  By introducing the supermode basis $a_{\pm}=(a_1 \pm a_2)/\sqrt{2}$, we can diagonalize $\Hopt$ which yields for the total Hamiltonian,

\begin{align}
\begin{split}
\Htot &= \Hnot + \Hint\, , \\
\Hnot &= \hbar \omega_{+}(0) \hat{a}_{+}^{\dagger} \hat{a}_{+} + \hbar \omega_{-}(0) \hat{a}_{-}^{\dagger} \hat{a}_{-} + \hbar \sum_{k} \omegamk \hat{b}_k^{\dagger} \hat{b}_k\, ,\\ 
\Hint &= \hbar \sum_{k} \xzpfk (\hat{b}_k +\hat{b}_k^{\dagger} ) \Big[ \frac{g_{1,k}+g_{2,k}}{2} ( \hat{a}_{+}^{\dagger} \hat{a}_{+} 
+ \hat{a}_{-}^{\dagger} \hat{a}_{-} ) + \frac{g_{1,k}-g_{2,k}}{2}(\hat{a}_{+}^{\dagger} \hat{a}_{-}+\hat{a}_{-}^{\dagger} \hat{a}_{+})\Big]\, ,
\end{split}
\label{eq:interaction}
\end{align}   

\noindent where the frequency difference between the supermodes at zero mechanical displacement is  $\omega_{+}(0)-\omega_{-}(0)= 2J$.  The first term inside the brackets of Eq.~(\ref{eq:interaction}) describes the \emph{linear self-mode} optomechanical coupling of the $a_{\pm}$ supermodes to the mechanical mode of interest. The last term describes the \emph{linear cross-mode}  optomechanical coupling, \emph{i.e.} the coupling between the $a_{\pm}$ supermodes mediated by the mechanical vibrations.  In the $a_{\pm}$ basis we have for the \emph{linear self-mode} optomechanical coupling to mechanical mode $b_{k}$,

\begin{equation}
g_{+,k}=g_{-,k}=\frac{g_{1,k}+g_{2,k}}{2}, 
\label{eq:gsupermodes}
\end{equation}

\noindent and the \emph{linear cross-mode} optomechanical coupling, 

\begin{equation}
g_{+-,k}=\frac{g_{1,k}-g_{2,k}}{2}.
\label{eq:gthreemode}
\end{equation}

Following the approach of Ref.~\cite{ludwig2012}, we further diagonalize the full Hamiltonian assuming a quasi-static approximation for the mechanical motion.  The resulting eigenfrequencies of the $a_{\pm}(\{\xk\})$ supermodes are,

\begin{equation}
\omega_{\pm}(\{\xk\})= \omega_0 + \sum_{k}g_{\pm,k}\xk \pm \sqrt{J^2+\left(\sum_{k}g_{+-,k}\xk \right)^2}\, ,
\label{eq:dispersion}
\end{equation} 

\noindent where the mechanical displacements ${x}_k=x_{\textnormal{zpf},k}(\hat{b}^{\dagger}_k+ \hat{b}_k)$ are regarded as a quasi-static variables. Focusing on a single mechanical mode $b_{k}$ and Taylor expanding the optical supermode frequencies as a function of small displacement $x_k$ around equilibrium position $\xkeq$ yields,
 
\begin{equation}
\omega_{\pm}(\{\xkeq+\xk\}) = \omega_\pm(\{\xkeq\}) + \delta\omega_{\pm,k}^{(1)}(\{\xkeq\})\xk + \delta\omega_{\pm,k}^{(2)}(\{\xkeq\})\xk^2 + ... \ \  ,
\label{eq:freq:expansion}
\end{equation}

\noindent where 

\begin{equation}
\delta \omega_{\pm,k}^{(1)}(\{\xkeq\}) = {\frac{\partial \omega_{\pm}}{\partial \xk} \bigg{|}}_{\{\xkeq\}} \equiv \g_{\pm,k}(\{\xkeq\})
\label{eq:first_order}
\end{equation}

\noindent and 

\begin{equation}
\delta \omega_{\pm,k}^{(2)}(\{\xkeq\}) = {\frac{1}{2} \frac{\partial^2 \omega_{\pm,k}}{\partial\xk^2}\bigg{|}}_{\{\xkeq\}} \equiv \gsq_{\pm,k}(\{\xkeq\}).
\label{eq:second_order}
\end{equation}

\noindent Here $\g_{\pm,k}(\{\xkeq\})$ is the linear (self-mode) coupling coefficient and $\gsq_{\pm,k}(\{\xkeq\})$ is the quadratic coupling coefficient of the $a_{\pm}(\{\xk\})$ supermode to the $k^{\text{th}}$ mechanical mode $b_{k}$.  In what follows we will be primarily interested in the linear and quadratic coupling coefficients around the symmetric equilibrium position $\{\xkeq\}=\{0\}$, where the linear coupling is trivially $\g_{\pm,k}(\{0\}) = \g_{\pm,k} = (g_{1,k}+g_{2,k})/2$ and the quadratic coupling can be related to the linear cross-mode coupling of the $a_{\pm}$ supermodes,

\begin{equation}
\gsq_{\pm,k}(\{0\}) \equiv \gsq_{\pm,k} = \pm\left({g}_{+-,k}\right)^2/2J\, .
\label{eq:gquadratic}
\end{equation}

\noindent We return to the more general result for $\xkeq \neq \{0\}$ in Sec.~\ref{subsec:couplingeq}.

From the above expressions, mechanical modes such that $g_{1,k}=-g_{2,k}$ will have vanishing \emph{linear self-mode} couplings to the $a_{\pm}$ supermodes, while mechanical modes such that $g_{1,k}=g_{2,k}$ will have vanishing \emph{linear cross-mode} coupling strength. All intermediate cases such as $|g_{1,k}| \neq |g_{2,k}|$ can of course occur in general.  In the following, we introduce a photonic crystal optomechanical resonator supporting multiple mechanical modes of different symmetries and with optimized overlap with the optical modes. We show that by engineering the bandstructure and defect of the photonic crystal the splitting $2J$ between the optical supermodes can be tuned to arbitrarily small values, which greatly enhances the $x^2$-coupling strength. 

\section{Multimode Photonic Crystal Optomechanical Cavity}
\label{sec:design}

\begin{figure}[t!]
\centering
\includegraphics[width=0.6\columnwidth]{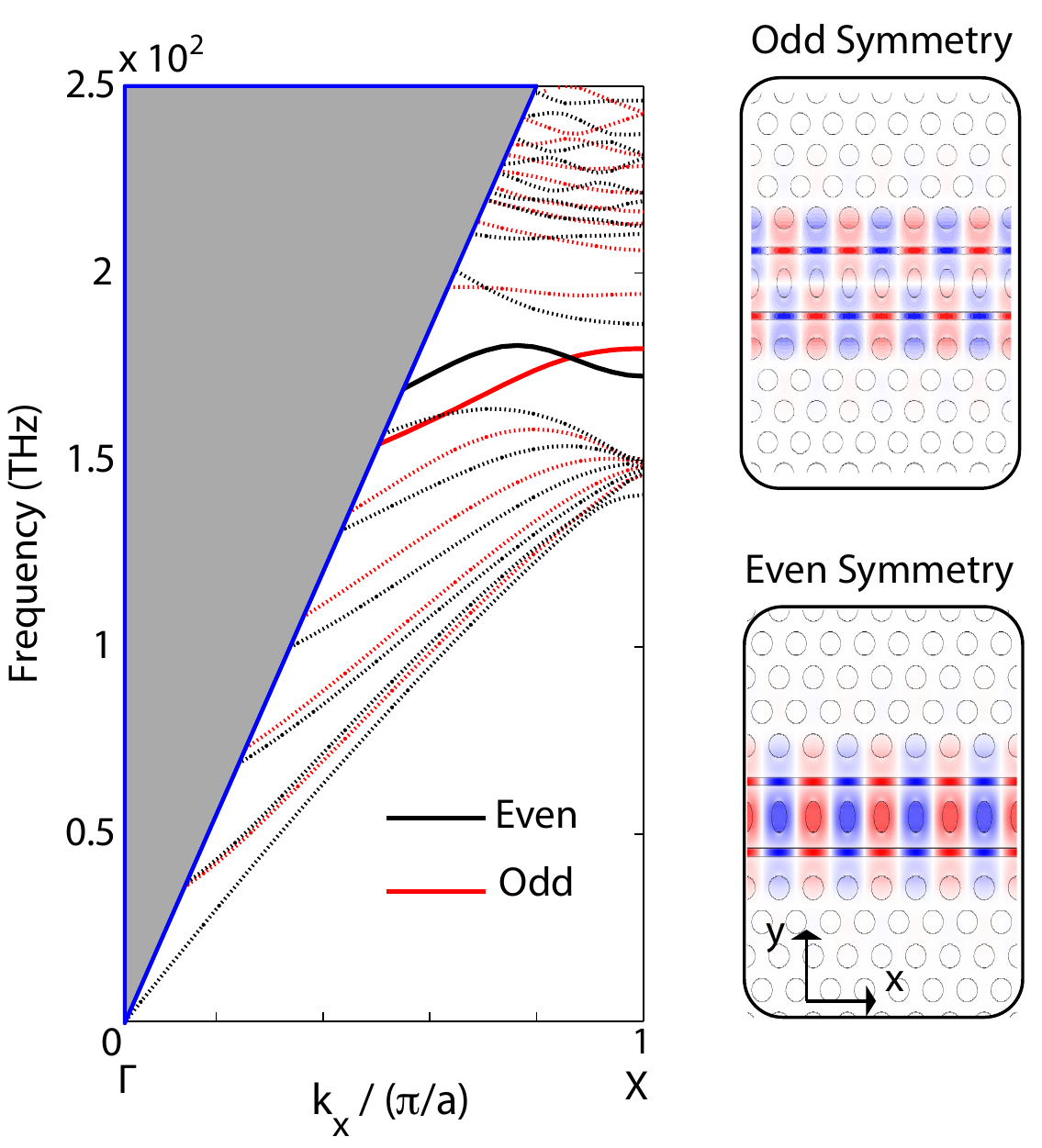} 
\caption{Bandstructure of a triangular lattice of air holes in a silicon slab with two line defects (air slots). We plot the TE-like bands with even (vector) symmetry about the slab mid-plane. The blue line and gray shaded region delimit the light cone of the air cladding surrounding the silicon slab. We focus on the fundamental waveguide modes (solid lines) inside the pseudo-bandgap of the triangular lattice. The $y$-polarization of the electric field, $E_y$, of the odd and even modes at the $X$-point are shown in the insets to the right of the bandstructure. Red and blue correspond to the positive and negative normalized amplitude of the $y$-polarized electric field.  The bandstructure here is computed for a silicon slab with the following set of parameters: refractive index $n=3.42$, thickness $t=220$~nm, lattice constant $a=480$~nm, hole radius $r=0.3 a$ and slot widths $s_1=s_2=100$~nm.}
\label{fig:bandprofile}
\end{figure}

\subsection{Bandstructure properties}
\label{sec:bandstructure}
In this paper, we design our structure assuming a silicon thin film device layer of thickness $t= 220$~nm, Young's modulus $Y=169$~GPa, mass density $\rho=2329$~kg$/$m$^3$ and refractive index $n=3.42$.  Our initial geometry is a quasi-two-dimensional (quasi-2D) periodic photonic crystal membrane patterned with a triangular lattice of holes (lattice constant $a=480$~nm,  circular hole radius $r=0.3a$)  This structure has an in-plane photonic bandgap (a pseudo-bandgap) for guided slab modes of predominantly TE polarization (electric field polarized in the plane of the slab) around a free-space wavelength of $\lambda \approx 1550$~nm.  Referring to Fig.~\ref{fig:design}(b), addition of the slots breaks the translational periodicity in the transverse direction ($y$-direction), leaving a periodic structure of lattice constant $a$ in the longitudinal ($x$) direction. Introducing two air slots of width $s_1=s_2=s \approx 100$~nm into the photonic crystal membrane results in a pair of optical waveguides with guided modes localized to each of the individual air slots due to the pseudo-bandgap.  As shown in Fig.~\ref{fig:design}(b), this splits the triangular lattice into two outer slabs and a central beam. 

\begin{figure}[t!]
\includegraphics[width=\columnwidth]{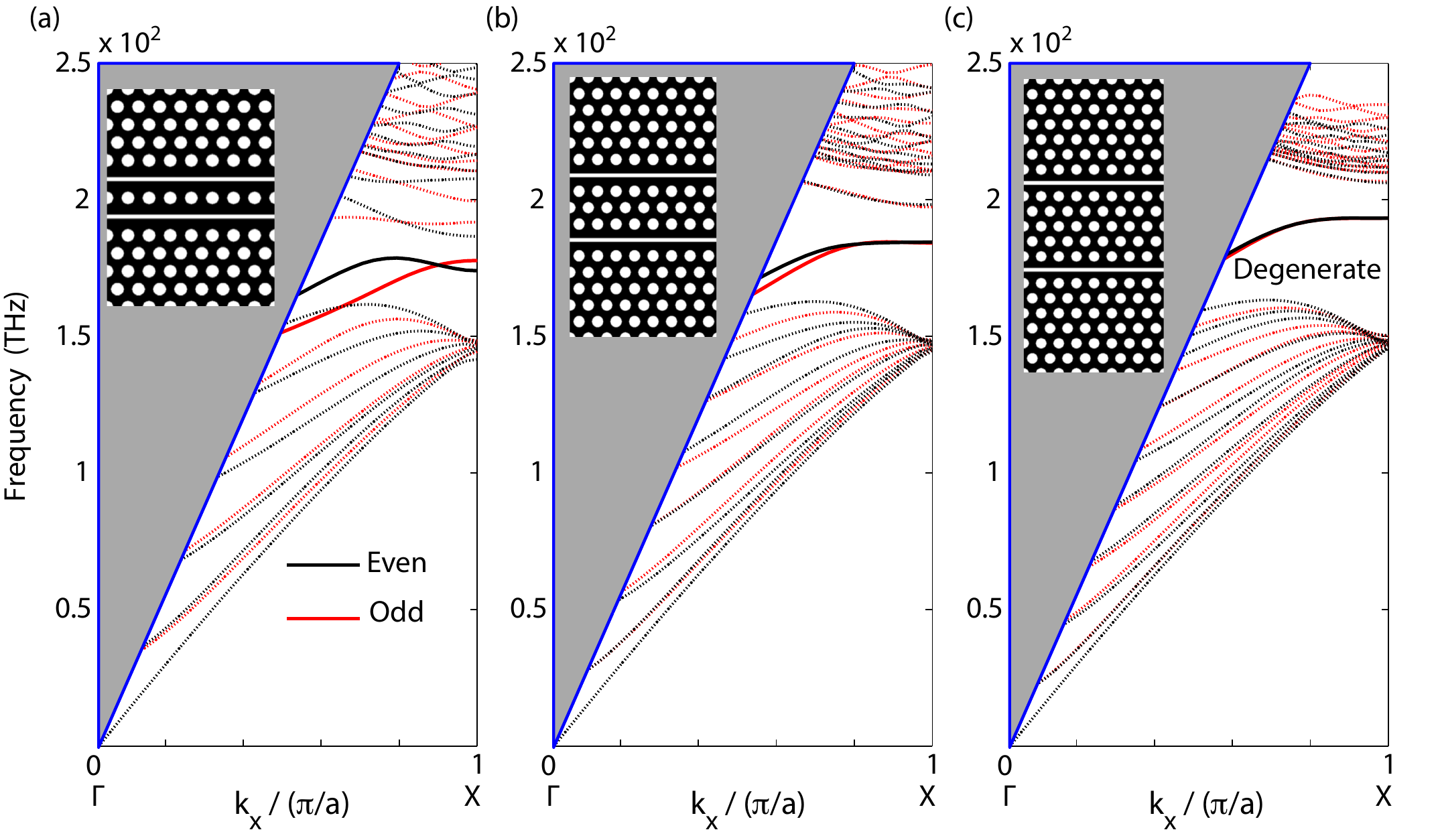}
\caption{Simulated bandstructures of the coupled linear waveguide modes for three different separations of the line defects: (a) one, (b) three and (c) five rows of holes in the central nanobeam. The waveguide modes of interest are shown as solid red (odd modes) and black (even modes) lines. The figure shows the decrease in frequency splitting between the targeted optical modes at the $X$-point as we increase the number of rows in the central nanobeam. The band diagrams are calculated for the same geometrical parameter as in Fig. \ref{fig:bandprofile}.}\label{fig:midsweep}
\end{figure}

\begin{figure}[t!]
\centering
\includegraphics[width=0.55\columnwidth]{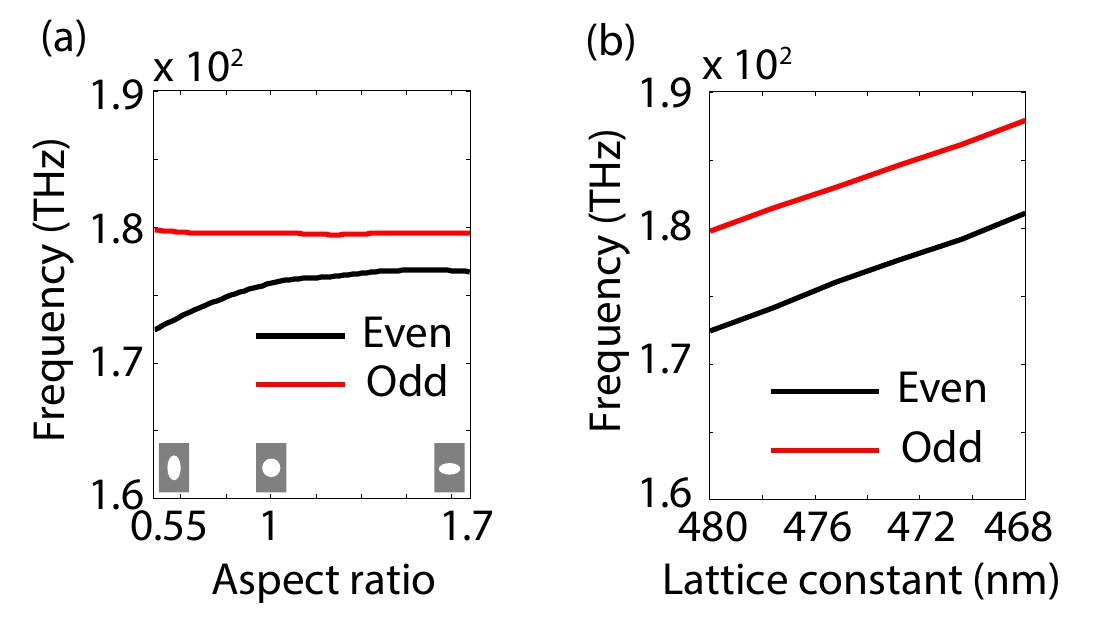} 
\caption{Influence of the hole ellipticity and lattice constant on the $X$-point bandedge frequency of the even and odd waveguide supermodes in the case of a single row of holes in the central nanobeam. (a) Shift of the $X$-point frequencies due to a change of hole ellipticity in the central nanobeam. The odd mode is unaffected by the change. The three insets illustrate the shape of holes for three different aspect ratios $\eta=0.55$, $1$ and $1.7$ with $(a,r,s_1,s_2)=(480$~nm$, 0.3a, 100$~nm$, 100$~nm$)$. (b) Increase of the optical waveguide supermodes' $X$-point frequency with a decrease of the lattice constant from $a=480$~nm to $a_D=468$~nm $= 0.975 a$ calculated with $(r,\eta,s_1,s_2) = (0.3a, 0.55, 100$~nm$, 100$~nm$)$.}
\label{fig:bandstructuredefects}
\end{figure}

The optical bandstructure of the guided modes of the double-slotted waveguide structure with a central beam consisting of a single row of air holes is plotted in Fig.~\ref{fig:bandprofile}.  The optical bandstructures here and in what follows are computed using the MIT Photonic Bands package~\cite{mpb}.  To simplify the bandstructure we only plot the optical bands with TE-like polarization (more accurately we plot those modes with even vector symmetry about the mid-plane of the photonic crystal thin-film slab).  In addition to the symmetry about the mid-plane of the silicon thin-film slab -- corresponding to the $\sigma_{z}$ mirror operator -- the double-slotted waveguide structure also has a symmetry plane about the $y=0$ plane as indicated by the green line in Fig.~\ref{fig:design}(b).  The mirror operator corresponding to this symmetry we label $\sigma_{y}$, and the modes of the double-slotted photonic crystal waveguide can be categorized by their even and odd parity under $\sigma_{y}$.  Here we use the labeling convention that the waveguide supermode with even $E_y$ electric field profile is called the ``even'' mode, while the waveguide supermode with odd $E_y$ electric field profile is termed the ``odd'' mode (note that classifying the modes by their vector symmetry would swap the mode labels).  The even and odd waveguide bands are shown as solid black and red lines in Fig.~\ref{fig:bandprofile}.  The back and red dashed curves correspond to the unguided (in the transverse $y$-direction) modes of the triangular photonic crystal slabs surrounding each air slot.  The shaded grey region corresponds to the region above the light cone of the air cladding surrounding the silicon slab, in which a continuum of radiation modes freely propagate out of the plane of the slab ($z$ direction).  The insets to the right of the bandstructure plot in Fig.~\ref{fig:bandprofile} show the $E_y$ field profiles of the even and odd waveguide modes at the edge of the first Brillouin zone ($X$-point). 

In the hopes of achieving large $x^2$-coupling strengths, our focus will be to design photonic crystal cavity modes with minimal optical splitting $2J$ (see Eq.~(\ref{eq:gquadratic})).  An obvious way to minimize the optical coupling between the slot waveguide modes would be to increase the separation between the slots.  This decreases the photon tunneling rate $J$, and hence decreases the frequency splitting between the waveguide modes. For instance, we show in Fig.~\ref{fig:midsweep} that the frequency splitting between the odd and even waveguide modes at the $X$-point can be decreased from $\sim3$~THz in the case of waveguide slots separated by a single row of holes, all the way down to $\sim68$~GHz for waveguide slots separated by five rows of holes. In the latter case the bands become nearly degenerate over a significant fraction of the Brillouin zone, whereas in the more strongly-coupled case of a single row of holes the even and odd waveguide bands have different slopes and even cross near the $X$-point. The mode profiles of the even and odd waveguide supermodes are plotted in Fig.~\ref{fig:bandprofile} for a separation between slots of a single row of holes, with the odd waveguide modes having a node in the center of the central beam, and thus, more of their energy in the air slots.  As described below, for a cavity based upon the coupled waveguides, this feature allows one to control the relative mode frequencies of the even and odd cavity supermodes by changing the size of the air slot gaps in the structure.

In addition to slot width and slot separation, the fact that the even waveguide modes have more of their energy in the central beam may also be exploited to tailor the relative frequencies of the waveguide supermodes.  We consider a middle slab consisting of a single row of holes (nanobeam) and analyze the impact of the ellipticity and lattice constant on the frequency of the waveguide modes. We parametrize the ellipticity of the holes by the aspect ratio $ \eta= {r_y/r_x}$, where $r_y$ ($r_x$) are the semi-axis of the ellipse in $y$ ($x$) direction, so that $\eta=1$ corresponds to circular holes. Further, by setting $r_y= r_0 /\sqrt{\eta}$ and $r_x = r_0 \sqrt{\eta}$ where $r$ is circular radius of the unperturbed cells, we keep the air filling fraction invariant between the elliptical holes and the circular holes. Fig.~\ref{fig:bandstructuredefects}(a) shows the frequency shift of the bands at the $X$-point due to the variation in the aspect ratio of the holes in nanobeam. Only the even band is influenced while the odd band does not shift. For completeness we show in Fig.~\ref{fig:bandstructuredefects}(b) the scaling of the even and odd waveguide modes at the $X$-point versus a scaling of the in-plane lattice constant (i.e., slab thickness constant).  As one might expect, the splitting between the waveguide supermodes of different parity are relatively unaffected by the lattice scaling. 

\begin{figure}[t!]
 \includegraphics[width=\columnwidth]{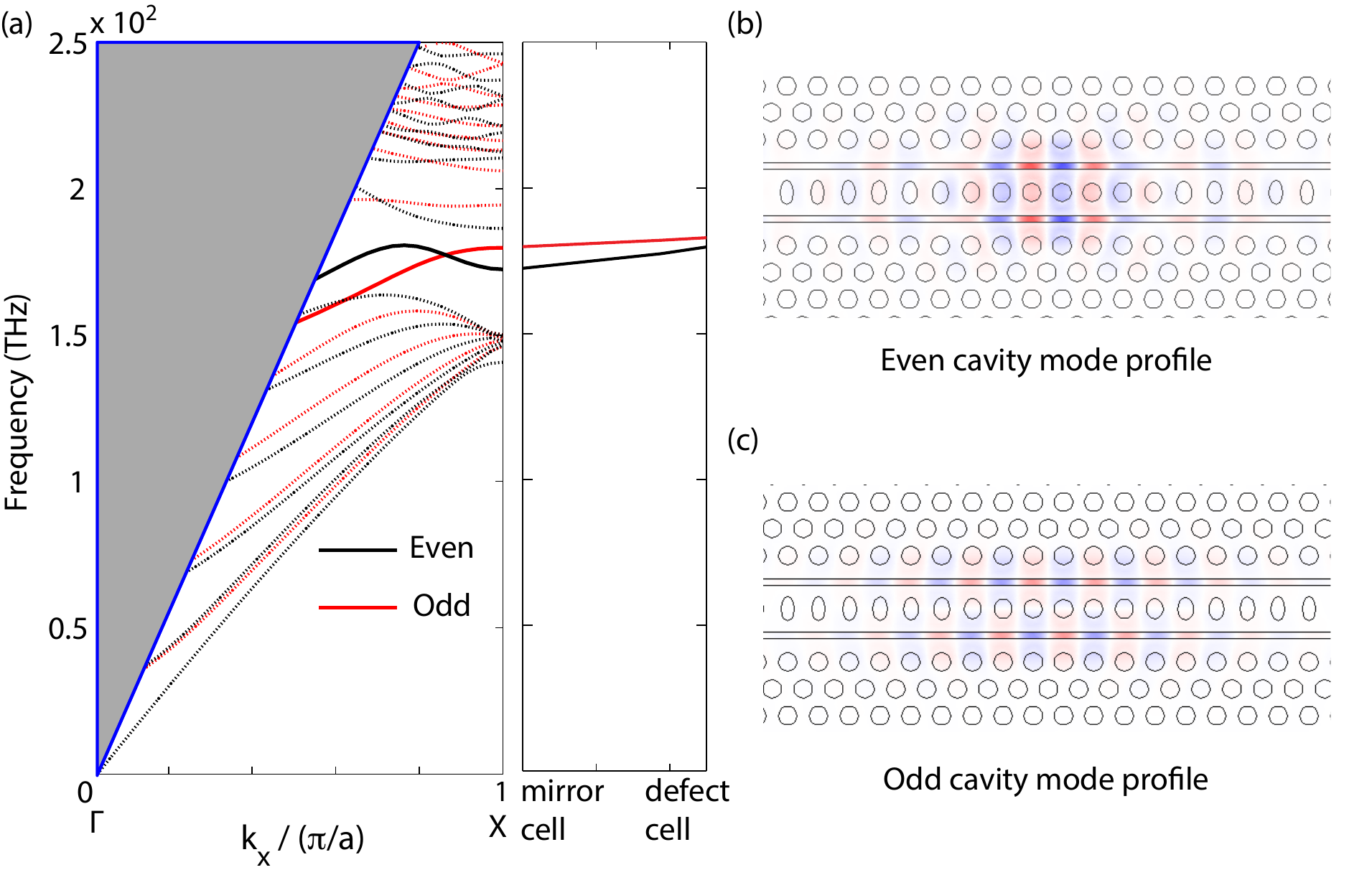} 
\caption{(a) Left: optical bandstructure of the mirror section of the cavity structure.  Right: waveguide supermode frequencies at the $X$-point as the waveguide unit cell transitions from the outer mirror section to the center of the defect region.  As described in the main text, this transition involves a change in the lattice constant ($a$) and in the nanobeam hole ellipticity ($\eta$). In the mirror cell, we use a lattice constant $a=480$~nm and elliptical holes with major axis of $r_y=194$~nm and minor axis of $r_x=107$~nm ($\eta = 0.55$). In the defect cell, we use $a=0.975 a$ and $\eta= 1$ (circular holes). The slot widths are $s_1=s_2$=100~nm.   The cavity is designed with total of $N_x=42$ waveguide unit cells along the $x$-axis, with a central defect region consisting of $N_D=7$ defect cells.  The outer slabs contain $N_y=9$ rows of holes in the transverse $y$-direction.  (b) Plot of the FEM-simulated amplitude of the $y$-polarization of the electric field $E_y(\vecb{r})$ of the even cavity mode.  (c)  Plot of the FEM-simulated odd cavity mode.  In (b) and (c), red and blue correspond to positive and negative $E_y$ field amplitudes, respectively.}
\label{fig:defect}
\end{figure}

\subsection{Optical cavity}
\label{sec:opticalcavity}
To form an optical cavity from the coupled waveguide system described above we need to find a way of closing the ends of the waveguides. To this end, we introduce a ``defect'' in the waveguide structure by modifying the geometrical parameters of the waveguide along its propagation ($x$) axis.  This defect region of the waveguide is then embedded between two ``mirror'' sections of the waveguide as depicted in Fig.~\ref{fig:design}.  Localized cavity modes result for a waveguide modification that pushes the $X$-point waveguide supermode frequencies in the defect unit cells inside the pseudo-bandgap of the unperturbed mirror unit cells~\cite{eichenfield2009picogram,chan2008optical}.

Here the cavity is designed by combining the two defects shown in Fig.~\ref{fig:bandstructuredefects}.  The lattice constant is decreased from $a=480$~nm to $a_D= 468$~nm quadratically over $N_D= 7$ cells at the center of the structure. Simultaneously with the scaling of the lattice constant, the aspect ratio of the air holes in the central nanobeam are increased from $\eta = 0.55$ to circular holes with $\eta =1$ while keeping a constant air filling fraction.  Fig.~\ref{fig:defect}(a) shows the modification of the waveguide mode frequencies from the unperturbed mirror cells to the central defect cell. The change in the lattice constant pushes the mode frequencies inside the bandgap and the change of hole ellipticity reduces the frequency splitting between the even and odd supermodes. We simulated the full cavity structure using the COMSOL Multiphysics~\cite{comsol} finite-element-method (FEM) solver. The length of the entire cavity consists of a total of $N_x=42$ waveguide unit cells along the $x$-axis, with $N_D= 7$ central defect unit cells.  Each of the outer photonic crystal slabs have $N_y=9$ rows of air holes in the transverse $y$-direction. Fig.~\ref{fig:defect}(b) and Fig.~\ref{fig:defect}(c) show the $E_y$ field profiles for the odd and the even cavity supermodes, respectively. With the given parameters, the odd mode has a simulated free-space wavelength of $\lambda=1539$~nm and a radiation-limited optical $Q$-factor of $Q=4.2 \times 10^6$.  The even mode has a free-space wavelength of $\lambda=1539.8$~nm and a substantially lower radiation-limited $Q$-factor of $Q=4 \times 10^5$.  This asymmetry in $Q$-factor results from both the larger coupling to even waveguide modes away from the $X$-point in the mirror section, a result of the non-monotonic dispersion of the even band, and the reduced vertical radiation loss for a mode of odd in-plane symmetry~\cite{Srinivasan2002}.  

\subsection{Dependence of the cavity supermode frequency splitting on the slot size}
\label{sec:crossing}

\begin{figure}[t!]
\centering
\includegraphics[width=0.75\columnwidth]{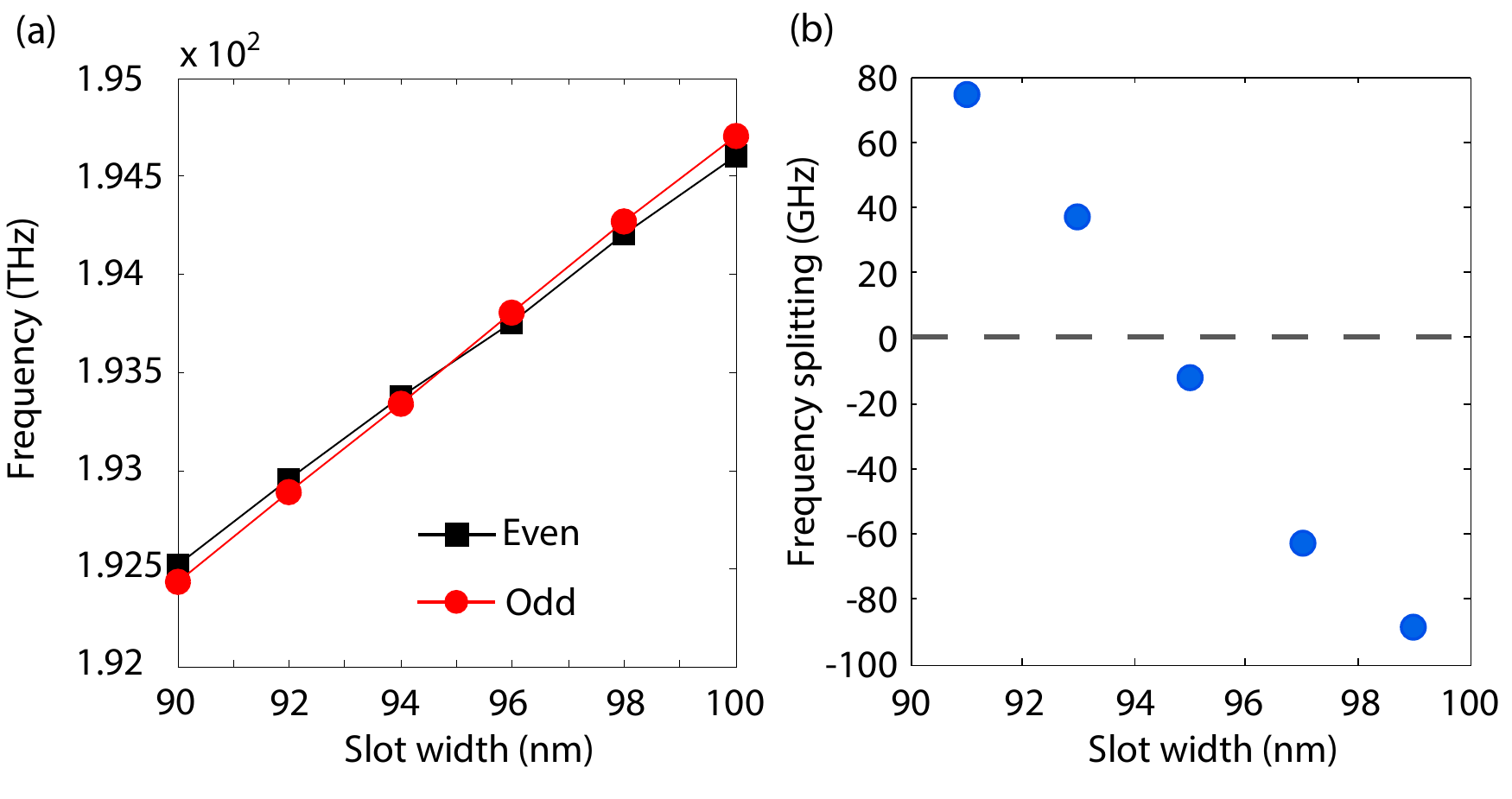}
\caption{(a) Cavity eigenfrequencies for a slot width varying from $s=90$~nm to $s=100$~nm. The frequency of the odd mode is more influenced by a change of the slot width than the even mode, leading to a change in the spectral ordering of the odd and even eigenmodes. (b) Extracted frequency splitting $\Delta \omega = 2J$ as function of the slot width. Arbitrarily small splittings are achieved around $s=95$~nm. The parameters of the structure are identical to Fig. \ref{fig:defect}.}
\label{fig:crossing}
\end{figure}

Since the optical modes of interest are mostly localized in the air slots, their frequencies are strongly impacted by the width of the air slots $s$. Increasing $s$ causes the effective refractive index of the optical waveguide modes to decreases, yielding a blue shift of the resulting optical cavity frequencies.  FEM simulations of the waveguide supermodes of the double-slotted photonic crystal structure for slot widths ranging from $s=90$~nm to $s=100$~nm are plotted in Fig. \ref{fig:crossing}~(a). The frequencies increase approximately linearly with $s$ with slightly different slopes, resulting in a crossing of the cavity supermodes at $s\sim95$~nm.  This crossing is made more apparent in Fig.~\ref{fig:crossing}(b), where the relative splitting between the even ($\omega_{+}$) and odd ($\omega_{-}$) supermode frequencies, $\Delta\omega=\omega_+-\omega_-=2J$, is plotted.

As we see, by solely modifying the slots width, it is possible to change the splitting between the cavity supermodes ($\Delta\omega=\omega_+-\omega_-$) from positive to negative. This is again explained by looking at the even and odd field profiles shown in Fig.~\ref{fig:defect}(b) and \ref{fig:defect}(c). Since the odd mode has a node in the middle of the central nanobeam, its effective refractive index is more sensitive to a change in the air region than the even cavity supermode. Therefore, for an equal change of the slot widths the frequency of the odd cavity supermode shifts more than the frequency of the even cavity supermode. It is also worth noting that the crossing of the \emph{cavity} supermodes is only possible because the odd and even \emph{waveguide} bands cross (see Fig.~\ref{fig:defect}(a)), with the even waveguide band having a \emph{higher} frequency than the odd waveguide band slightly away from the $X$-point.  

\begin{figure}[t!]
\centering
\includegraphics[width=0.6\columnwidth]{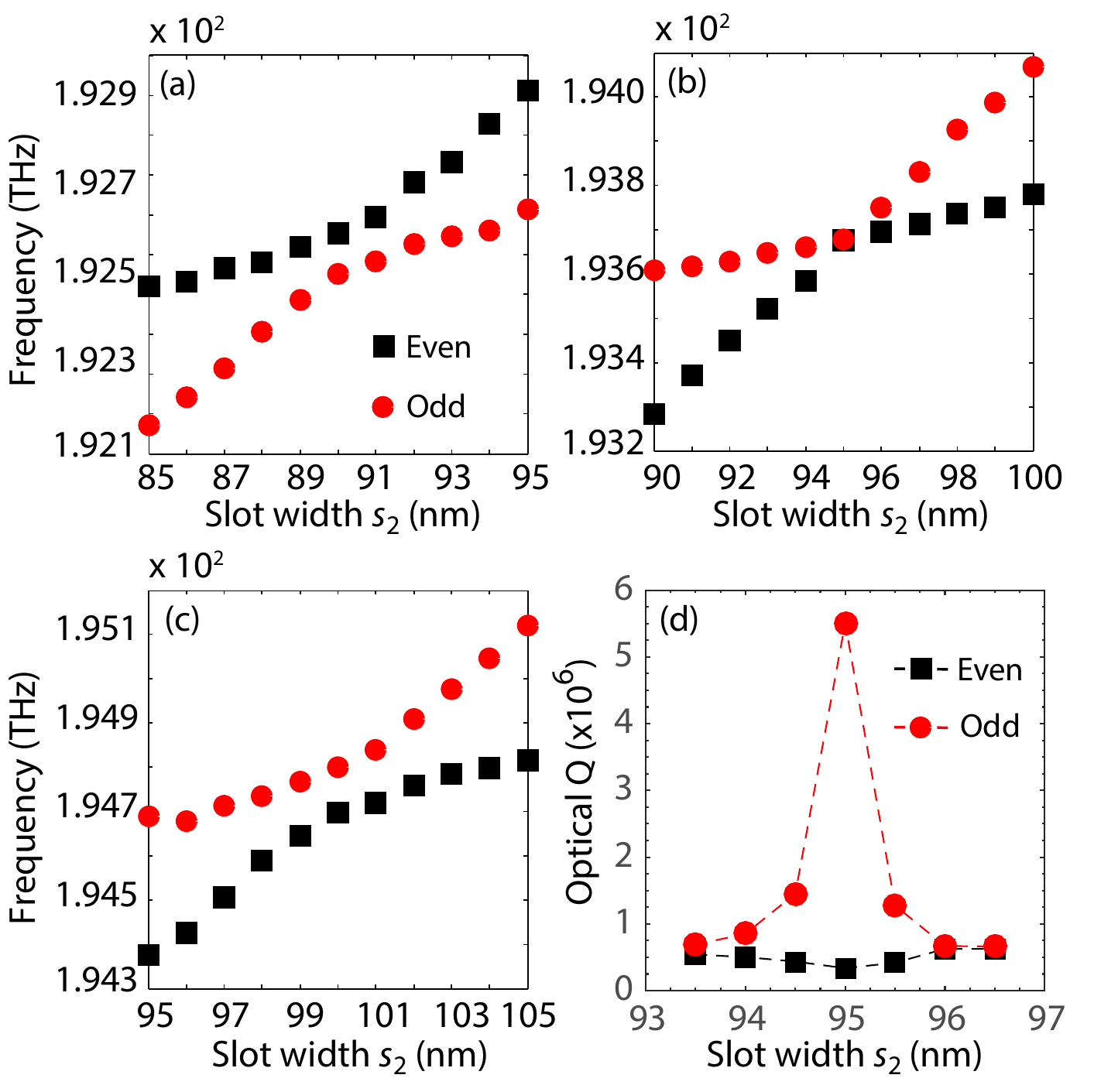}
\caption{Simulated anti-crossing curves obtained by varying $s_2$ while $s_1$ is kept constant at (a) $90$~nm (b) $95$~nm and (c) $100$~nm.  The splitting between the odd (red circles) and even (black squares) cavity supermodes is inverted between (a) and (c). In (b), the splitting is reduced from $2J/2\pi>100$~GHz to $2J/2\pi=17$~GHz. (d) Plot of the $Q$-factor of the even and odd supermode branches versus the second slot width $s_{2}$ for fixed slot width $s_{1}=95$~nm.  The parameters of the structure, save the slot width, are the same as in Fig.~\ref{fig:defect} for all simulations in (a-d).}
\label{fig:anticrossing}
\end{figure}

Figure \ref{fig:anticrossing}(a-c) shows anti-crossing curves obtained where one slot width ($s_1$) is fixed and the other is swept from $s_2=s_1-5$~nm to $s_2=s_1+5$~nm.  In this case, the cavity mode eigenfrequencies anti-cross with a minimal splitting $2J$ achieved for the symmetric $s_2=s_1$ situation.  The odd and even supermode branches of the anti-crossing curves were identified in Fig.~\ref{fig:anticrossing} by looking at the parity of electric field profile of the simulated eigenmodes at the center of the anti-crossing (here we label the branches by the parity of the supermodes at the center of the anti-crossing curve).  Notice that the lower frequency branch for slot width $s_1=90$~nm (Fig.~\ref{fig:anticrossing}(a)) is the odd cavity supermode, whereas the higher frequency branch for slot width $s_1=100$~nm (Fig.~\ref{fig:anticrossing}(c)) has odd parity.  For the in-between case of $s_1=95$~nm, the splitting at the center of the anti-crossing curve is approximately zero ($2J/2\pi=17$~GHz), as it must be if the parity of the upper and lower frequency supermode branches swap.  The possibility to control the supermode frequency splitting in this way provides a unique way to tune the strength of the $x^2$-coupling according to Eq.~(\ref{eq:gquadratic}).

In addition to tuning the cavity supermode splitting, adjusting the air slot sizes may also be used to tune the absolute and relative optical $Q$-factor of the two cavity supermodes.  Figure~\ref{fig:anticrossing}(d) plots the simulated radiation-limited $Q$-factor for the two cavity supermodes as the slot widths are adjusted and the cavity modes sweep through the anti-crossing point.  At the center of the anti-crossing curve the two cavity supermodes are to a good approximation even and odd parity modes around the $y=0$ plane, which as we pointed out earlier results in a significant difference in their $Q$-factor.  Motion of the outer photonic crystal slabs or the central nanobeam will also change the relative air slot sizes, thus changing the radiation damping of the optical cavity supermodes.  As noted in recent theoretical work~\cite{Yanay2016}, this \emph{dissipative} optomechanical coupling can lead to interference of the quantum noise entering the cavity mode system.  This in turn can be used to cool the coupled mechanical resonator to the quantum ground-state even in the unresolved sideband regime (bad-cavity limit).  More relevantly, in the case of a predominantly $x^2$ coupled system, this effect can be used to reduce the parasitic linear back-action and enable continuous $x^2$ measurements of the mechanical motion.           

%\begin{figure}[t!]
%\centering
%\includegraphics[scale=0.6]{Figure7b.pdf}
%\caption{Simulated anti-crossing curves obtained by varying $s_2$ while $s_1$ is kept constant at (a) 90~nm (b) 95~nm and (c) 100~nm.  The splitting between the odd (red circles) and even (black squares) branches is inverted between (a) and (c). In (b), the splitting is reduced from $2J/2\pi>100$~GHz to $2J/2\pi=17$~GHz. (d) Dependence of the even and odd $Q$ factors as a function of $s_2$ extracted from the simulations of (b). The parameters of the structure are the same as in Fig. \ref{fig:defect}.}\label{fig:anticrossing}
%\end{figure}

% Fig 7b (d)
%The optical quality factors also found to strongly depend on the position along the anti-crossing. In Fig. \ref{fig:anticrossing} (d), we display the $Q$ factors of the normal modes in the case where  $s_1=95$~nm and $s_2$ is swept from 92~nm to 97~nm. The $Q$ factor of the odd mode increases from $7\times10^5$ away from the anti-crossing to $6\times10^6$ at the anti-crossing while the $Q$ factor of the even mode decreases from $5\times 10^5$ away from the anti-crossing to $ 3\times 10^5$ at the anti-crossing. The even mode has a lower quality factor because the non-monotonic decrease of the even band away from the $X$ point favors the scattering of the defect mode to non-localized waveguide modes of different wavevectors. 

\section{Mechanical Resonances}
\label{sec:mechanicaldesign}

Having considered the optical cavity modes of the double-slotted planar photonic crystal, we now analyze the mechanical modes of the structure.  In order to support mechanical resonances the photonic crystal slabs are suspended.  The optical modes can interact with both flexural  and localized acoustic modes of the central nanobeam.  In Sec.~\ref{sec:bulkmechanics}, we present the flexural modes of the structure. We show that higher orders flexural modes are found to exist up to 1~GHz with significant optomechanical coupling, making them suitable for operation in the resolved sideband regime, where the optical linewidth $\kappa$ is much smaller than the mechanical frequency $\omegam$ \cite{RevModPhys.86.1391}. In Sec.~\ref{sec:mechanicalcavity}, we show that the defect developed to form the optical cavity also gives rise to a localized acoustic resonance of a few GHz frequency. 

\subsection{Flexural mechanical resonances}\label{sec:bulkmechanics}
\begin{figure}[t!]
\centering
\includegraphics[width=\columnwidth]{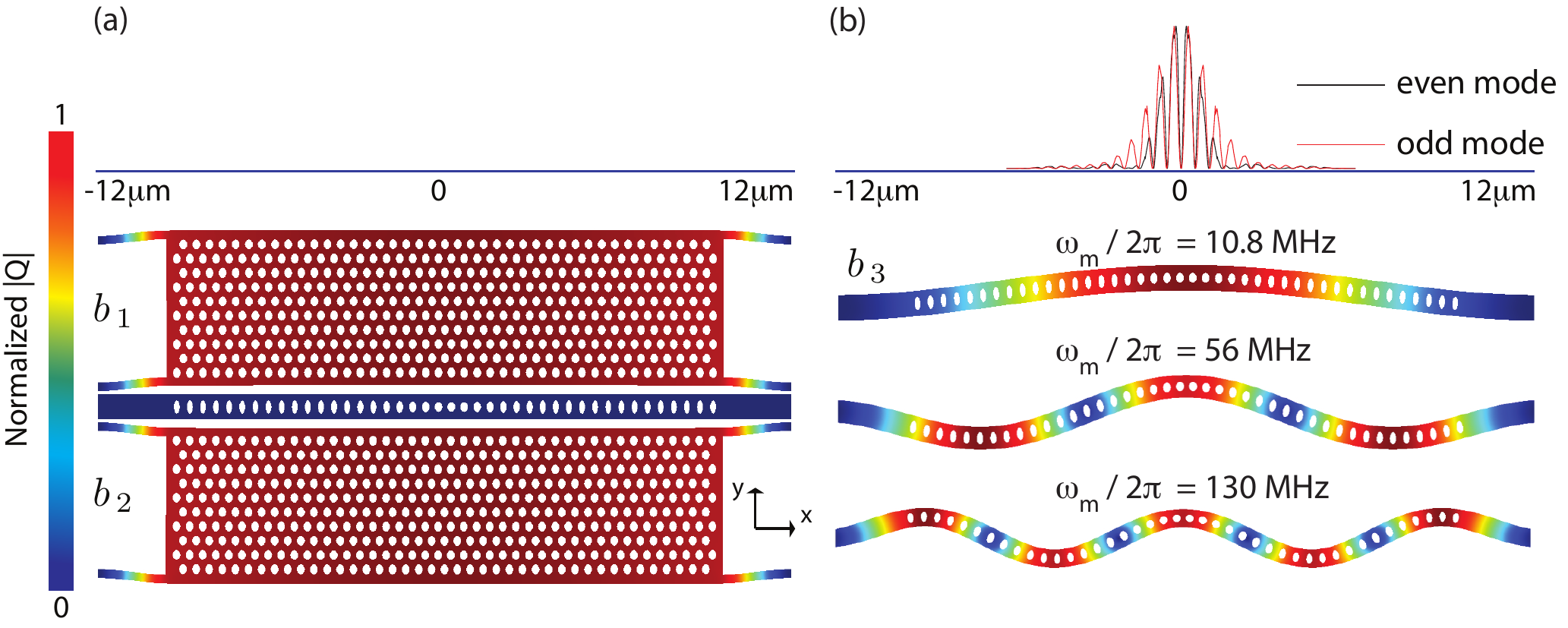}
\caption{Normalized displacement profile of  (a) the in-plane slab modes and (b) the nanobeam first and higher order in-plane flexural modes. 
The inset on top of (b) shows the profiles of $|E_y|^2$ along the waveguides for both the odd and even symmetry optical supermodes. The deformations are exaggerated for clarity. The photonic crystal parameters are the same as in Fig. \ref{fig:defect}. The central nanobeam is $731$~nm wide and 24~$\mu$m long. The outer slabs are suspended by tethers of length $l_t=2.5$~$\mu$m and width $w_t=150$ nm. }\label{fig:mechbulkmodes}
\end{figure}

The outer slabs and nanobeam behave as three independent mechanical resonators supporting various in-plane and out-of-plane flexural mechanical resonances. Here, we focus on the in-plane flexural modes that are asymmetric with respect to the $y=0$ plane mirror operator ($\sigma_y=-1$), and symmetric with respect to the $x=0$ plane mirror operator ($\sigma_x=+1$), about the center of the structure. These flexural modes are represented in Fig.~\ref{fig:mechbulkmodes} with exaggerated deformation profiles. Since our main focus will be on the fundamental resonances, we denote the fundamental in-plane flexural modes of the two outer slabs and nanobeam as $b_1$, $b_2$ and $b_3$ respectively.  Their respective frequencies are denoted $\omega_{b_1}$, $\omega_{b_2}$ and $\omega_{b_3}$.

In our design, the outer slabs are suspended by tethers of length  $l_t=$2.5~$\mu$m and width $w_t=$150~nm, yielding fundamental in-plane flexural resonance of $\omegam/2\pi\simeq 6$~MHz. As shown in Fig.~\ref{fig:mechbulkmodes}~(a), these modes correspond to a uniform displacement of the whole slabs. The displacement of one outer slab causes a uniform change of the width of the adjacent slot, and hence a change of the optical supermode frequencies. The in-plane slab modes provide degrees of freedom for the electromechanical tuning of the slot widths. 

In Fig.~\ref{fig:mechbulkmodes}~(b), we plot the displacement profiles of the first three lowest frequency ($10.8$~MHz, $56$~MHz and $130$~MHz) nanobeam in-plane flexural modes of symmetry $\{\sigma_x=+1,\sigma_y=-1\}$. The $y$-polarized electric field profiles $|E_y|^2$ are plotted in the inset for both the odd and even optical supermodes $a_{\pm}$. The finite extent of the optical modes along the $x$-axis of the photonic crystal structure limits the region of the nanobeam that will contribute to the optomechanical interaction. As a result, the nanobeam displacement amplitude $x_3$ can be approximated by a net effective displacement of the whole nanobeam $\bar{x}_3 \approx x_3$, causing one slot width to change by an amount $+\bar{x}_3$ and the other to change by $- \bar{x}_3$. Because of this asymmetric displacement, the optomechanical couplings of these flexural mode to the individual slot modes $a_1$ and $a_2$ are expected to be equal and of opposite sign. This favors the quadratic and linear cross-mode interaction terms introduced in Eqs.~(\ref{eq:gthreemode}) and (\ref{eq:gquadratic}).

Higher order flexural modes of the nanobeam have been identified with frequencies up to $1$~GHz and are summarized in table~\ref{table:nbmodes}. Assuming a moderate optical quality factor of $Q=5\times 10^5$, the resolved-sideband regime condition ($\kappa<\omegam$) could be met with a flexural mode of frequency $\omegam/2\pi=400$~MHz.

\begin{table}[b]
\begin{center}\caption{\small In-plane flexural modes of the nanobeam. We consider the modes with symmetric displacement with respect to $\sigma_x$. The geometric parameters of the nanobeam are the same as in Fig. \ref{fig:defect}.} 
\begin{tabular}{l*{11}{c}r}
\hline 
$\omegam/2\pi$   [MHz]     &  10.8 & 56  & 130 & 227 & 340 & 467 & 605 & 746 & 884  & 1025 \\
$\xzpf$   [fm] & 15.8 & 6.6 & 4.2 & 3.1 & 2.5 & 2.2 & 1.9 &  1.7  & 1.6 & 1.5  \\
$\meff$   [pg]  & 3.1 & 3.4 & 3.6 & 3.9 & 3.9 & 3.8 & 3.8 & 3.8  &  3.8 & 3.6 \\
\hline 
\end{tabular}
\label{table:nbmodes}
\end{center}
\end{table}

\subsection{Localized phononic crystal resonance}
\label{sec:mechanicalcavity}

In Sec.~\ref{sec:opticalcavity} we described how to localize optical waveguide modes of the double-slotted photonic crystal waveguide propagating by engineering a perturbation to the waveguide unit cell in the propagation direction.  In particular, we analyzed the photonic bandstructure of the waveguide unit cell and designed a defect based on a combination of change in the lattice constant and change in the central nanobeam hole aspect ratio. Here we study the phononic bandstructure of the nanobeam unit cell and show that our choice of photonic crystal defect parameters makes the nanobeam compatible with the localization of an GHz-frequency acoustic resonance.

\begin{figure}[t!]
\centering
\includegraphics[width=\columnwidth]{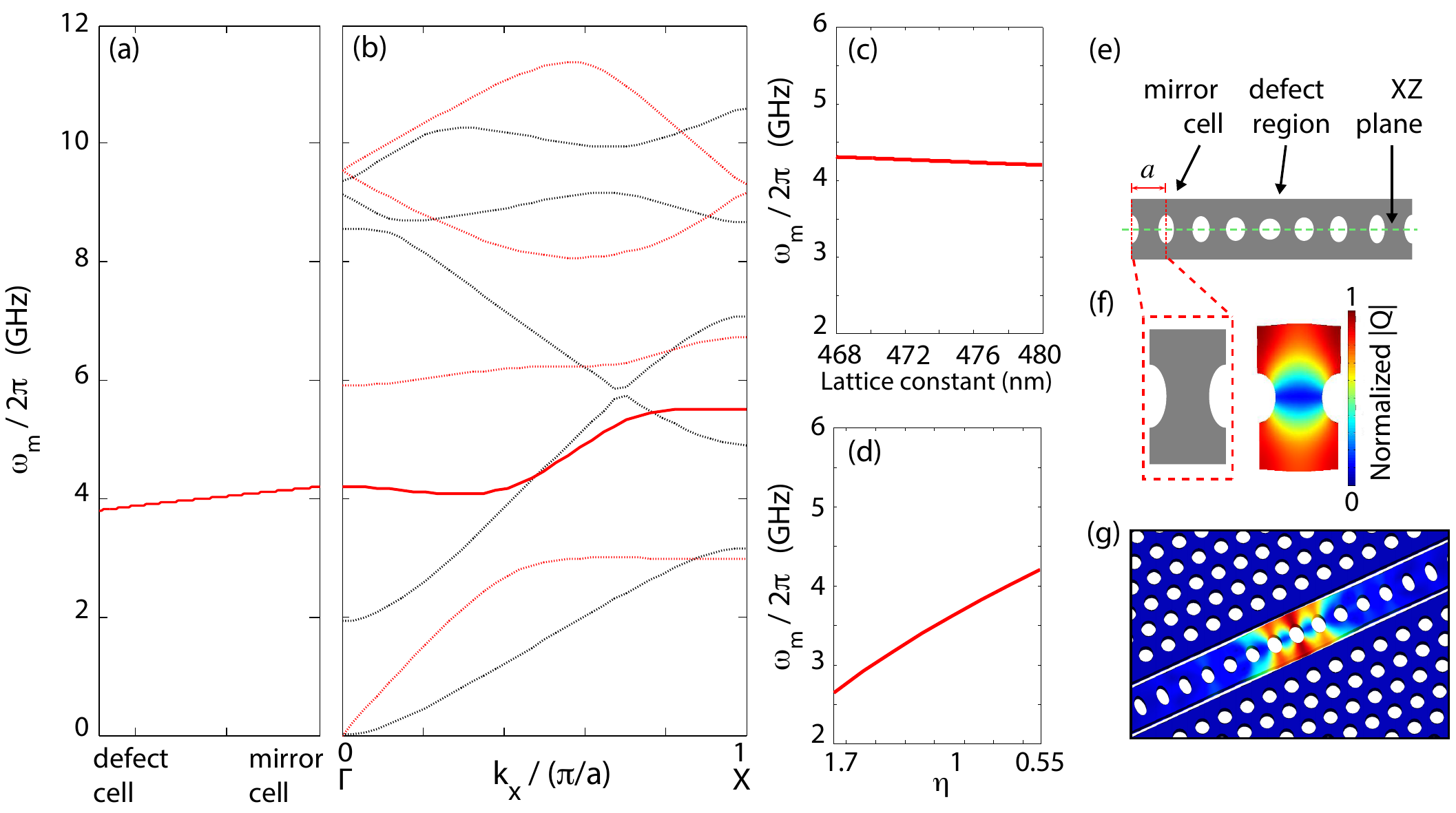} 
\caption{(a,b) Simulated phononic bandstructure of the nanobeam and defect mode drawn from the $\Gamma$-point. The breathing mode band is specified by the solid red line. The even (red lines) and odd (black lines) symmetry acoustic modes are defined with respect to $\sigma_y$ mirror operator.  (a) Shift of the $\Gamma$-point frequency of the breathing mode band for a defect formed by both variations in lattice constant and ellipticity of holes. The defect parameters are the same as in Fig. \ref{fig:defect}. (c) and (d) show the shift of the breathing mode frequency at the $\Gamma$ point due to variations of the lattice constant and of the holes aspect ratio, respectively. (e)  Mechanical beam and (f) normalized displacement field $\bf{Q}(\bf{r})$ of the localized breathing mode. The color scale indicates the magnitude of $\bf{Q}(\bf{r})$. (g) Exaggerated deformation of the structure due to the breathing mode.  All acoustic mode simulations were performed using COMSOL~\cite{comsol}.} \label{fig:mechband}
\end{figure}

Figure~\ref{fig:mechband}(a) and \ref{fig:mechband}(b) show the FEM-simulated acoustic bandstructure of the nanobeam unit cell and the frequency shift of the \emph{breathing} mode band at the $\Gamma$-point as the nanobeam transitions from the mirror unit cell geometry to the defect unit cell geometry. The breathing mode band is shown as a solid red curve. The nanobeam unit cell and the corresponding normalized displacement field profile $\bf{Q}(\bf{r})$ of the breathing mode are depicted in Figs.~`\ref{fig:mechband}(e-g).  The localized breathing mode is drawn from the $\Gamma$-point of the bandstructure in order to have a significant optomechanical coupling to the optical mode~\cite{mattopto2009}.

Figure~\ref{fig:mechband}(c) and \ref{fig:mechband}(d) detail separately the shifts of $\Gamma$-point frequency of the breathing mode caused by the perturbation in the lattice constant and hole ellipticity of the nanobeam. Increasing the aspect ratio $\eta$ will push the breathing mode frequency down into the band gap of the unperturbed mirror cells, while decreasing the lattice constant slightly increases the frequency of the $\Gamma$-point mode.  As summarized in Fig. \ref{fig:mechband}(a), the same defect used to localize the photonic crystal resonances satisfies the conditions to localize a phononic crystal resonance. Note that, in contrast to the mechanical flexural modes, the displacement of the localized breathing mode is symmetric with respect to $\sigma_y$ so the optomechanical coupling to this mode is expected to be the same for both $a_1$ and $a_2$. Therefore, the breathing mode is expected to have negligible linear cross-mode and quadratic coupling strengths. Nevertheless, the breathing mode presents the advantage of being well in the resolved sideband regime and could potentially be used as an auxiliary mechanical mode in multimode optomechanically induced transparency schemes as proposed in Refs.~\cite{Fan-Tang-cascaded-eit,Ojanen2014}.

\section{Optomechanical Coupling Relations}
\label{sec:coupling}

With knowledge of the optical and mechanical properties of the double slotted waveguide cavity, we can now turn to the calculation of the different optomechanical coupling factors.  We utilize a perturbation theory of Maxwell's equations~\cite{johnsonperturbation} suitable for dealing with both spatial shifts in the dielectric boundaries of the cavity structure as well as stress-induced modifications of the local dielectric constant of the deformed structure.  Applying first-order perturbation theory to the numerically computed unperturbed optical field profiles and mechanical field profiles allows us to evaluate both the linear self-mode coupling $g_{\pm}$ and the linear cross-mode coupling $g_{+-}$ for the nanobeam flexural modes~\cite{braginsky2002analysis,miao,zhao2009}.  By considering the perturbation theory to second-order yields the strength of the quadratic coupling in our structure.  Finally, we consider the modification of the coupling strengths due to deviations of the structure from the symmetric equilibrium position of the central nanobeam and outer slabs.

\subsection{Linear self-mode optomechanical coupling}
\label{subsec:linearcoupling}
Maxwell's equations in a source-free, linear dielectric medium, yields the following eigenvalue equation for the electric field,

\begin{equation}
\nabla \times \nabla \times \ket{E} = \left( \frac{\omega^2}{c}\right) \epsilon \ket{E},
\label{eq:Eeigen}
\end{equation}

\noindent where we have used the Dirac notation $\ket{E}$ for the electric field $\vecb{E}(\vecb{r})$ eigenstate with harmonic time dependence $e^{-i\omega t}$.  Here, $c$ is the speed of light in vacuum, and $\epsilon(\vecb{r})$ is a dielectric constant which is a function of the spatial coordinate $\vecb{r}$ (and most generally a tensor). We are interested in the change in the modal frequency due to an infinitesimal perturbation $\delta\alpha$ to the dielectric structure. The first-order correction term to the mode frequency is expressed as

\begin{equation}
\delta\omega^{(1)} = -\frac{\omega^{(0)}}{2}\frac{\bra{E^{(0)}} \delta\alpha \ket{E^{(0)}} }{\bra{E^{(0)}}\epsilon^{(0)} \ket{E^{(0)}}},
\label{eq:delta_omega1}
\end{equation}

\noindent where $\epsilon^{(0)}(\vecb{r})$ is the unperturbed dielectric constant of the structure, $\ket{E^{(0)}}$ and $\omega^{(0)}$ are electric field and frequency of the harmonic optical mode of interest, $\bra{E^{(0)}}\epsilon^{(0)} \ket{E^{(0)}} \equiv \int_{V}\text{d}^3r\left(\vecb{E}^{(0)}\cdot \left(\epsilon^{(0)}\vecb{E}^{(0)}\right)^{\ast}\right)$, and $\delta\alpha$ depends upon the type of perturbation to the dielectric structure. 

The change in dielectric constant due to mechanical displacement arises from two main contributions.  The first contribution is due to shifting of the interface boundary between two dielectric media.  In this case the dielectric function is a high-contrast step function which translates displacements normal to the boundary into local modifications of the dielectric seen by the electric field.  As proposed by Johnson, et al., in Ref.~\cite{johnsonperturbation}, the appropriate perturbation theory in this so-called \emph{moving boundary} (MB) problem is given by,

\begin{equation}
\delta\omega^{(1)}_{\text{MB}} = -\frac{\omega^{(0)}}{2} \frac{\int_{A} \mathrm{d}^2r (\mathbf{q}\cdot\mathbf{n})\ [ \Delta \epsilon |E_{\|}^{(0)}|^2 - \Delta \epsilon^{-1} |D_{\bot}^{(0)}|^2]}{\bra{E^{(0)}}\epsilon^{(0)} \ket{E^{(0)}}},
\label{eq:MB}
\end{equation}

\noindent where $\epsilon_{1(2)}$ is the dielectric constant of medium 1 (2) at any point in the boundary surface $A$ between two media of differing dielectric constant, $\Delta \epsilon = \epsilon_{1} - \epsilon_{2}$, $\Delta \epsilon^{-1}= \epsilon^{-1}_{1} - \epsilon_2^{-1}$, $|E_{\|}^{(0)}|$ ($|D_{\bot}^{(0)}|$) is the magnitude of the unperturbed electric (displacement) field polarized in the plane (out of the plane) of the boundary surface $A$ between medium 1 and medium 2, and $\mathbf{n}(\vecb{r})$ is the outward unit vector normal pointing from medium 1 into medium 2 on boundary $A$. Here, $\mathbf{q}(\mathbf{r})$ is the normalized displacement field of the mechanical mode of interest with maximum displacement equal to unity, $\max|\mathbf{q}(\mathbf{r})| = 1$.  We can also define an effective mass of the mechanical mode in terms of $\vecb{q}$, 

\begin{equation}
\meff = \int_V \mathrm{d}^3r \rho(\vecb{r})|\mathbf{q}(\mathbf{r})|^2,
\label{eq:effectivemass}
\end{equation}

\noindent where $\rho$ is the mass density of the dielectric material defining the optomechanical structure.  This effective mass is the appropriate motional mass for evaluating the zero-point fluctuation amplitude, $\xzpf = \sqrt{{\hbar}/{2\meff\omegam}}\ $, of the generalized amplitude coordinate corresponding to the point of maximum amplitude of the mechanical mode.  

The second contribution to the linear self-mode coupling is due to the \emph{photoelastic effect}, resulting from the change of the dielectric constant due to the local strain induced by the mechanical displacement.  The first-order perturbation to the dielectric tensor is given by,

\begin{equation}
\delta\pmb{\epsilon}= - \pmb{\epsilon}^{(0)} \frac{\mathbf{p S}}{\epsilon_{0}} \pmb{\epsilon}^{(0)},
\label{eq:deltaeps_PE_tens}
\end{equation}

\noindent where $\pmb{\epsilon}^{(0)}$ is the unperturbed dielectric tensor, $\epsilon_{0}$ is the permittivity of free space, $\mathbf{p}$ is the fourth rank photoelastic tensor, and $\mathbf{S}$ is the symmetric strain tensor. For an isotropic medium this simplifies to,
 
\begin{equation}
\delta\epsilon_{ij} = - \epsilon_0 n^4 p_{ijkl} S_{kl},
\label{eq:deltaeps_PE}
\end{equation} 

\noindent in index notation.  In matrix form,

\begin{align}
\begin{split}
&\delta\pmb{\epsilon} = -\epsilon_0 n^4  \\
&\times \left[
    \begin{array}{ccc}
      p_{11}S_{xx}+p_{12}(S_{yy}+S_{zz}) & p_{44}S_{xy} & p_{44}S_{xz} \\
      p_{44}S_{xy} & p_{11}S_{yy}+p_{12}(S_{xx}+S_{zz}) &p_{44}S_{yz}\\
      p_{44}S_{xz} &  p_{44}S_{yz} &p_{11}S_{zz} + p_{12}(S_{xx}+S_{yy})\\
    \end{array}
  \right]. 
\end{split}
\label{eq:PEdepsilon}
\end{align}  

\noindent The resulting first-order photoelastic (PE) correction to the optical frequency is

\begin{multline}
%\begin{split}
\delta\omega^{(1)}_{\text{PE}}  = \frac{\omega^{(0)} \epsilon_0 n^4}{2 \bra{E^{(0)}} \pmb{\epsilon}^{(0)} \ket{E^{(0)}}} \int_{V} \mathrm{d}^3r \,2\Big[\mathbf{Re}((E^{(0)}_x)^{\ast}E^{(0)}_y)p_{44}S_{xy}\\
+  \mathbf{Re}((E^{(0)}_x)^{\ast}E^{(0)}_z)p_{44}S_{xz} + \mathbf{Re}((E^{(0)}_y)^{\ast}E^{(0)}_z)p_{44}S_{yz} \\
+  |E^{(0)}_x|^{2}(p_{11}S_{xx}+p_{12}(S_{yy}+S_{zz})) +  |E^{(0)}_y|^{2}(p_{11}S_{yy}+p_{12}(S_{xx}+S_{zz})) \\
+  |E^{(0)}_z|^{2}(p_{11}S_{zz}+p_{12}(S_{yy}+S_{xx}))\Big].
%\end{split}
\label{eq:PEdomega}
\end{multline}

\noindent In the structures studied here, which are made by etching patterns into a thin-film of silicon, the only two media are silicon and vacuum.  As such, for the PE contribution to the linear self-mode coupling we utilize the photoelastic tensor coefficients for silicon in evaluating the integral in the numerator~\cite{biegelsen1975frequency}: $p_{11}=-0.0101$, $p_{12}=0.009$ and $p_{44}=-0.051$.

We begin by considering the calculation of $g_{+,b_1}$ and $g_{-,b_1}$ for our double-slotted photonic crystal device, i.e., the linear optomechanical couplings of the optical supermodes $a_+$ and $a_-$ to the fundamental in-plane flexural mode of either outer slab (we choose $b_1$ in this case). Table~\ref{table:anticrossing} displays the numerically computed coefficients using the perturbation theory described above in terms of the unperturbed optical and mechanical fields.  $g_{+,b_1}$ and $g_{-,b_1}$ can also be approximated by fitting the anti-crossing curves of Fig.~\ref{fig:anticrossing} using the dispersion relation given in Eq.~(\ref{eq:dispersion}).  The approximate dispersion relation fit values for the linear couplings are also shown in Tab.~\ref{table:anticrossing}, and compare well to the exact perturbation theory values despite the fact that the couplings derived from the dispersion relations using Eqs.~(\ref{eq:gsupermodes}) and (\ref{eq:dispersion}) neglect cross-coupling terms between $a_1$ and $a_2$ mediated by the mechanics.  

The localized breathing mode of the central nanobeam was found by FEM simulations to be at a mechanical frequency of $\omegam/2\pi\approx 4$~GHz, with linear coupling rates of $\tilde{g}_{+}/2\pi=$249~kHz and $\tilde{g}_{-}/2\pi=$163~kHz to the $a_{\pm}$ supermodes, respectively, where we have used the notation $\tilde{g}_{\pm}={g}_{\pm} x_{\textnormal{zpf}}$.

\begin{table}[t]
\centering\caption{\small Strength of the linear optomechanical coupling of the optical supermodes to the fundamental in-plane flexural modes of the \emph{outer slabs} for three different slot widths. The second and third columns display the linear coupling strengths calculated numerically using the perturbation theory. The fourth column gives the values of the optomechanical coupling constant obtained by fitting the anti-crossing curves shown in Fig.~\ref{fig:anticrossing}. The geometric parameters of the nanobeam are the same as in Fig.~\ref{fig:mechbulkmodes}.}
\begin{tabular}{c*{3}{c}}
\hline
 & \multicolumn{2}{c}{First order perturbation theory}& Anti-crossing fit \\
Slot width & ${g}_{+,b_1}/2\pi$ & ${g}_{-,b_1}/2\pi$& $[ (g_{1,b_1}+g_{2,b_1})/2]/2\pi$ \\
$[$nm$]$ & $[$GHz/nm$]$ & $[$GHz/nm$]$& $[$GHz/nm$]$ \\
\cline{1-4}
90 & 55.05 & 57.79 & 51.13 \\
95 & 51.81 & 53.62 & 48.81 \\
100 & 41.12 &47.69 & 44.45 \\
\hline
\end{tabular}
\label{table:anticrossing}
\end{table}

\subsection{Linear cross-mode optomechanical coupling}
\label{subsec:crossmode}
By analogy with Eq.~(\ref{eq:delta_omega1}), the first order perturbation term for the linear cross-mode coupling $g_{ij}$, where $i\neq j$, can be written as~\cite{chang2011slowing} 

\begin{equation}
{g}_{ij,k}= -\frac{\sqrt{\omega_{i}^{(0)}\omega_{j}^{(0)}}}{2} \frac{\bra{E_i^{(0)}} \delta\alpha_{k} \ket{E_j^{(0)}} }{\left(\bra{E_i^{(0)}}\epsilon^{(0)}\ket{E_i^{(0)}}\right)^{1/2}\left(\bra{E_j^{(0)}}\epsilon^{(0)}\ket{E_j^{(0)}}\right)^{1/2}} \ ,
\label{eq:pert:threemode}
\end{equation}

\noindent In the case of the double-slotted photonic crystal of this work, we have for the shifting boundaries contribution to the cross-mode coupling between the supermodes $a_+$ and $a_-$ at the symmetric ($\{\xkeq\}=0$) equilibrium position (center of the anti-crossing curve of Fig.~\ref{fig:anticrossing}):

\begin{equation}
{g}_{+-,k} = -\frac{\sqrt{\omega_{+}^{(0)}\omega_{-}^{(0)}}}{2} \frac{\int_{A} \text{d}^{2}r (\vecb{q}_{k} \cdot \vecb{n})\ [ \Delta \epsilon \left(E_{\|,+}^{(0)}\right)^{*} \cdot E_{\|,-}^{(0)} - \Delta \epsilon^{-1} \left(D_{\bot,+}^{(0)}\right)^{*} \cdot D_{\bot,-}^{(0)} ]}{\left(\bra{E_{+}^{(0)}}\epsilon^{(0)}\ket{E_{+}^{(0)}}\right)^{1/2}\left(\bra{E_{-}^{(0)}}\epsilon^{(0)}\ket{E_{-}^{(0)}}\right)^{1/2}} \ .
\label{eq:pert:threemode2}
\end{equation}

\noindent Note that for the flexural mechanical modes of the photonic crystal structure (either slab or central nanobeam modes) we expect this to be the dominant contribution to the optomechanical coupling.  Expanding $\vecb{E}_+=(\vecb{E}_1+\vecb{E}_2 )/\sqrt{2}$ and $\vecb{E}_-=(\vecb{E}_1-\vecb{E}_2 )/\sqrt{2}$ in terms of the slot modes, and neglecting the cross terms such as $\vecb{E}_{1}^{*} \cdot \vecb{E}_{2}$ due to the small spatial overlap between the fields of the modes localized in separated slots, we obtain Eq.~(\ref{eq:gthreemode}) again, $g_{+-,k}=(g_{1,k}-g_{2,k})/2$. 

Consider now the flexural modes of the central nanobeam. At the symmetric ($\{\xkeq\}=0$) equilibrium position, the nanobeam's in-plane flexural modes are such that $g_1=-g_2$. Therefore, at the center of the anti-crossing curve $g_{+-}$ is maximal and equal to the linear coupling of the nanobeam mode to the $a_{1,2}$ slot modes.  Table~\ref{table:parametric-quadratic} shows the numerically computed linear cross-mode coupling rate $\tilde{g}_{ij}=g_{ij}\xzpf$  for the fundamental and higher order nanobeam in-plane flexural modes along with their respective frequencies simulated for slot sizes $s_1=s_2=95$~nm using Eq.~(\ref{eq:pert:threemode2}). Due to the tight localization of optical modes (see Fig.~\ref{fig:mechbulkmodes}) we find there is still significant coupling to higher order flexural modes of frequencies all the way up to $1$~GHz. For the numerical simulations of the $a_{\pm}$ optical supermodes of the double-slotted photonic crystal structure we also find that the radiation-limited optical quality factor is theoretically equal to $5 \times 10^6$ and $3 \times 10^5$ for the odd and even modes respectively.  Therefore, as noted earlier, we can expect mechanical modes of frequencies $\omegam/2\pi>300$~MHz to be in the resolved-sideband regime.

\subsection{Quadratic optomechanical coupling}
\label{subsec:quadcoupl}
By extending the perturbation theory to the second order, it is also possible to calculate the $x^2$-coupling strength \cite{cohen-tannoudji,Kaviani2014-paddle,Rodriguez2011,Amir_thesis2013}. We obtain, for a given optical mode $a_i$

\begin{equation}
\delta\omega_i^{(2)} = \frac{3}{8}{{\omega^{(0)}}} \left | \frac{\bra{E_i^{(0)}} \delta\alpha \ket{E_i^{(0)}} }{\bra{E_i^{(0)}}\epsilon^{(0)} \ket{E_i^{(0)}}} \right |^2 - \frac{1}{2} \sum_{j\neq i}  \frac{{\omega_i^{(0)}}^3}{{\omega_j^{(0)}}^2 -{\omega_i^{(0)}}^2} \frac{\left | \bra{E_j^{(0)}} \delta\alpha \ket{E_i^{(0)}} \right |^2}{\bra{E_j^{(0)}}\epsilon^{(0)} \ket{E_j^{(0)}}\bra{E_i^{(0)}}\epsilon^{(0)} \ket{E_i^{(0)}}} \ .
\label{eq:pert:quadratic}
\end{equation} 

\noindent In the case of the supermodes $a_+ \, $ and $a_-$ of the symmetric double-slotted photonic crystal structure ($\{\xkeq\}=0$), the first term vanishes and the only contribution to the  $x^2$-coupling comes from the second term.  For optical splittings such that $2J \ll \omega_{0}$, 

\begin{align}
\begin{split}
\delta\omega_+^{(2)}(\{0\})  \equiv \gsq_+& \approx  - \frac{{\omega_+^{(0)}}}{({\omega_+^{(0)}} + {\omega_-^{(0)}})({\omega_-^{(0)}} -{\omega_+^{(0)}})} \frac{\left(\omega_+^{(0)}\right)^2\left | \bra{E_-^{(0)}} \delta\alpha \ket{E_+^{(0)}} \right |^2}{\bra{E_-^{(0)}}\epsilon^{(0)} \ket{E_-^{(0)}}\bra{E_+^{(0)}}\epsilon^{(0)} \ket{E_+^{(0)}}} \ , \\
& \approx   \frac{g_{+-}^2}{2J}    \ ,
\end{split}
\label{eq:pert:quadratic:approx}
\end{align} 

\noindent which is what we obtained in Eq.~(\ref{eq:gquadratic}).  In Eq.~(\ref{eq:pert:quadratic:approx}) we only consider the contribution from the fundamental optical cavity supermodes because the frequency splitting between them is relatively small. Note that another approach~\cite{Kaviani2014-paddle} has shown that using a large number of spatially overlapping optical modes rather than decreasing the splitting of just two optical modes (as in our case) can also lead to significant $x^2$-coupling strengths.  The values of $\gsqzpf=\gsq(\xzpf)^2$ are summarized in Tab.~\ref{table:parametric-quadratic} for the nanobeam in-plane flexural modes up to $884$~MHz. Here we assume an optical $a_{\pm}$ supermode frequency splitting of $2J/2\pi=1$~GHz, which is close to the minimum splitting based on the estimated optical quality factors which allows the optical supermodes to be selectively excited and interrogated.  

\begin{table}[h!]
\centering\caption{\small Linear cross-mode optomechanical (vacuum) coupling rates $\tilde{g}_{+-}$ of the optical supermodes to the nanobeam's in-plane fundamental and higher order flexural modes. The $x^2$-coupling rate $\gsqzpf_+$ is inferred using Eq.~(\ref{eq:gquadratic}) for a minimum splitting of $2J/2\pi=1$~GHz. The geometric parameters of the nanobeam are the same as in Fig. \ref{fig:mechbulkmodes}.}
\begin{tabular}{r*{3}{c}r}
\hline 
$\omegam/2\pi$ [MHz] & $\tilde{g}_{+-}/2\pi$ [kHz] & $\gsqzpf_{+}/2\pi$ [Hz]\\
\hline
10.8 & 1020 &  1000 \\
56   &  402   &  160   \\
130 &  271   &  73     \\
227 &  208   &  43     \\
340 &  167   &  28     \\
467 &  126   &  15     \\
605 &  81   &    7       \\
746 &   44   &   2       \\
884 &  20   &    0.4    \\
1025 &  8   &    0.07  \\
\hline 
\end{tabular}
\label{table:parametric-quadratic}
\end{table}

\subsection{Coupling coefficients as a function of a static displacement of the nanobeam}
\label{subsec:couplingeq}
In the analysis of the optomechanical coupling coefficients described above in Section~\ref{subsec:linearcoupling}-\ref{subsec:quadcoupl} we considered a symmetric double-slotted structure with equal slot widths $s_1=s_2$ (equilibrium position $\{\xkeq\}=0$), and thus the calculations were done for the optical supermodes $a_{\pm}$.  In practice, we could find this symmetric condition by tuning one of the slabs to adjust the relative slots sizes until the optical frequency spectrum was at the center of the anti-crossing curve as shown in Fig.~\ref{fig:anticrossing}.  Far away from the center of the anti-crossing, however, the optical supermodes correspond more closely to the individual slot modes $a_1$ and $a_2$, and we expect different optomechanical coupling strengths. Here we describe how the optomechanical coupling coefficients change upon a large, static displacement of the central nanobeam which takes us far from the symmetric condition near the center of the anti-crossing curve. 

From the approximate analytical expression of the supermode dispersion (see Eq.~\ref{eq:dispersion}), we derive here approximate expressions for $g_{\pm,b_3}(x_{3,\text{eq}})$, $g_{+-,b_3}(x_{3,\text{eq}})$ and $\gsq_{\pm,b_3}(x_{3,\text{eq}})$ as a function of the static displacement amplitude $x_{3,\text{eq}}$ of the fundamental nanobeam mode:

%\begin{eqnarray}
%g_{\pm,b_3}(x_3) & =& \frac{g_{1,b_3}+g_{2,b_3}}{2}\pm \frac{{\left(\frac{g_{1,b_3}-g_{2,b_3}}{2}\right)}^2 x_3}{\sqrt{J^2 + \left(\frac{g_{1,b_3}-g_{2,b_3}}{2}\right)^2{x_3}^2} } \label{eq:quasistat:lin} \\
%{g}_{+-,b_3}(x_3) & = &\frac{{\left(\frac{g_{1,b_3}-g_{2,b_3}}{2}\right) \cdot J}}{\sqrt{J^2 + \left(\frac{g_{1,b_3}-g_{2,b_3}}{2}\right)^2{x_3}^2} }  \label{eq:quasistat:threemode}\\
%g'_{\pm,b_3}(x_3) & = & \pm \ \, \frac{1}{2} \cdot \frac{{g_{+-,b_3}(0)^2 \cdot J^2}}{\left({J^2 + g_{+-,b_3}(0)^2{x_3}^2}\right)^{\frac{3}{2}} }\label{eq:quasistat:quad} 
%\end{eqnarray}
%where we explicitly indicate the relation between $g'(x)$ and $g_{+-}(0)=g_{\pm,b_3}(\infty)$.

\begin{align}
g_{\pm,b_3}(x_{3,\text{eq}}) & \approx \frac{g_{1,b_3}+g_{2,b_3}}{2}\pm \left(\frac{g_{1,b_3}-g_{2,b_3}}{2}\right)\frac{{Z}}{\sqrt{1+{Z}^2} }, \label{eq:quasistat:lin} \\
{g}_{+-,b_3}(x_{3,\text{eq}}) & \approx \left(\frac{g_{1,b_3}-g_{2,b_3}}{2}\right)  \frac{1}{\sqrt{1+{Z}^2} },  \label{eq:quasistat:threemode}\\
g'_{\pm,b_3}(x_{3,\text{eq}}) & \approx \pm \ \,  \frac{g_{+-,b_3}^2}{2J} \left[\frac{1}{\sqrt{1+{Z}^2}}\right]^{3}\label{eq:quasistat:quad} 
\end{align}

\noindent where ${Z} = \left(\left(g_{1,b_3}-g_{2,b_3}\right)/2J\right) x_{3,\text{eq}}$. Note $g_{+-,b_3} \equiv g_{+-,b_3}(x_{3,\text{eq}}=0)$ as per our previously established convention.

\begin{figure}[t]
\centering
\includegraphics[width=\columnwidth]{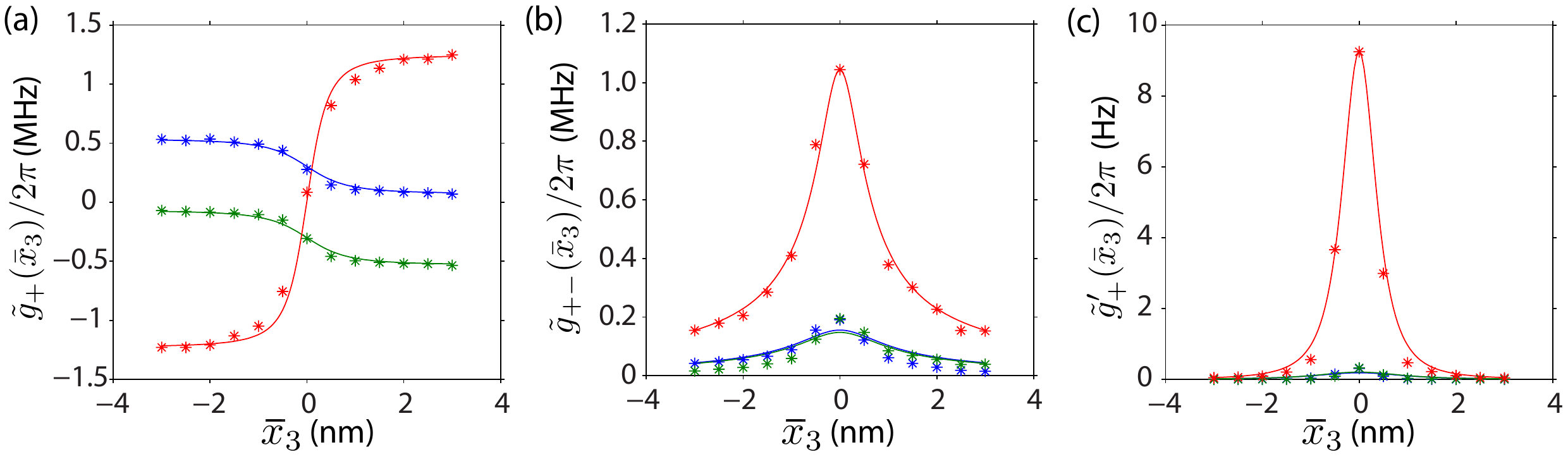}
\caption{Optomechanical coupling rates as a function of static central nanobeam displacement.  (a) Linear \emph{self-mode} optomechanical coupling, (b) linear \emph{cross-mode} optomechanical coupling, and (c) $x^2$-coupling of the $\omega_{+}(\bar{x}_{3})$ supermode branch to the fundamental in-plane mechanical resonances of the nanobeam ($b_{3}$, red curve) and outer slabs ($b_{1}$, blue curve; $b_{2}$, green curve).  The asterisk ($*$) correspond to numerical FEM simulations of the coupling rates using the perturbation theory with different in-plane static displacement $\bar{x}_3$ of the nanobeam from its symmetric equilibrium position.  The geometrical parameters of the simulated double-slotted structure are: $(a,r,s_1,s_2)=(480$~nm$, 0.3a, 90$~nm$+\bar{x}_3 , 90$~nm$-\bar{x}_3 )$. The solid lines are theoretical fits based on Eqs.~(\ref{eq:quasistat:lin}-\ref{eq:quasistat:quad}).}
\label{fig:quasistatic}
\end{figure}

As discussed in Sec.~\ref{sec:bulkmechanics}, $x_{3,\text{eq}}$ can be approximated by a static displacement $\bar{x}_3$ of the whole nanobeam. Using the perturbative calculation of the optomechanical coupling coefficients from the numerically simulated optical and mechanical fields one can then obtain $g_{\pm,b_3}(\bar{x}_3)$, $g_{+-,b_3}(\bar{x}_3)$ and $\gsq_{\pm,b_3}(\bar{x}_3)$ by simulating a structure with the nanobeam displaced from its equilibrium position by $\bar{x}_3$ (this becomes the new ``unperturbed'' structure in our perturbative calculations). Here we consider a structure with nominal slot widths of $s=90$~nm at equilibrium, and the displacement of the nanobeam from its equilibrium position is swept from $\bar{x}_3=-3$~nm to $\bar{x}_3 = 3$~nm in steps of $0.5$~nm. At each position we calculate the coupling coefficients between the optical supermodes and the fundamental in-plane flexural mode of the nanobeam.  The results of these simulations and calculations are plotted in Fig.~\ref{fig:quasistatic}. A fit to the numerically calculated coefficients using Eqs.~(\ref{eq:quasistat:lin}-\ref{eq:quasistat:quad}) show very good agreement. 

These results confirm our previous speculations that far from the symmetric equilibrium position the linear optomechanical coupling is the dominant optomechanical interaction between the optical supermodes and the flexural modes of the central nanobeam, while at symmetric equilibrium position of the beam ($\bar{x}_3 =0$)  the optomechanical interaction is predominantly $x^2$-coupling. The linear cross-mode coupling between the optical supermodes is also maximal at $\bar{x}_3 =0$, however its effects are strongly suppressed in the present case since the splitting between the optical resonances (here $2J/2\pi=117$~GHz) is much larger than the mechanical frequency of the fundamental in-plane nanobeam mode ($10.8$~MHz) \cite{Chen2015,Miao2009}.  Experimentally, a static displacement of the whole nanobeam by $\bar{x}_3$ can be mimicked by displacing both outer slabs in the same direction by an amount $\bar{x}_3$.  Conversely, a change in the equilibrium slot size can be achieved by displacing both outer slabs in opposite directions. In our recent experimental realization of the double-slotted photonic crystal cavity structure we used this tuning degree of freedom by integrating a set of independent capacitors on the outer slabs~\cite{Paraiso2015}.

\section{Summary}\label{sec:result}
We presented a general formalism for studying linear and quadratic coupling of mechanical motion with optical fields in a multimoded optical cavity.  This formalism is used to model and design a double-slotted photonic crystal cavity structure previously studied by us experimentally in Ref.~\cite{Paraiso2015}. The device supports two high-$Q$ optical resonances at telecommunication wavelengths and mechanical resonances spanning both the unresolved sideband and resolved sideband regimes. We find that depending on the symmetry of the photonic crystal structure, the optical supermodes can be made to interact either linearly or quadratically with mechanical motion. In particular, we show that the splitting between the optical supermodes can be strongly suppressed, which provides a significant enhancement of the quadratic ($x^2$) coupling. The linear and quadratic optomechanical coupling coefficients are calculated using perturbation theory and finite-element-method simulations of the (unperturbed) optical modes and mechanical displacement fields. The simulated optical quality factors of the photonic crystal cavity are calculated to be of order $Q \geq 5\times 10^5$, placing mechanical modes of frequencies $\omegam/2\pi\geq 400$~MHz in the resolved sideband regime. With such parameters, it is shown that the designed photonic crystal device can achieve zero-point $x^2$-couplings as large as $\gsqzpf/2\pi=1$~kHz for the $10.8$~MHz fundamental in-plane flexural mode of the structure, and $\gsqzpf/2\pi = 1$~Hz for a higher order flexural mode of frequency $884$~MHz in the resolved sideband regime, several orders of magnitude larger than in any other proposed optomechanical system to date.  Notably, the splitting of the photonic crystal cavity supermodes may be tuned by adjusting the cavity slot widths to a minimal value approaching that of the cavity mode linewidths ($\sim 1$~GHz), greatly enhancing the achievable $x^2$ coupling.  Although QND measurements of the stored mechanical energy will require substantially lower optical cavity losses~\cite{miao,ludwig2012,Paraiso2015}, with this scale of $x^2$-coupling it is feasible to consider a number of other interesting experiments with slightly less restrictive parameters.  These include the quantum measurement of phonon or photon shot noise~\cite{Clerk2010}, or utilizing interference of quantum noise in the bad-cavity limit~\cite{Yanay2016}, a continuous position measurement of $x^2$.  The latter measurement requires additional dissipative optomechanical coupling, which the photonic crystal cavity structure of this work also possesses.     

\begin{acknowledgments}
This work was supported by the AFOSR Hybrid Nanophotonics MURI (FA9550-12-1-0024), the Institute for Quantum Information and Matter, an NSF Physics Frontiers Center (NSF Grant PHY-1125565) with support of the Gordon and Betty Moore Foundation (GBMF-2644), the Alexander von Humboldt Foundation, and the Max Planck Society.
\end{acknowledgments}

%\section*{Acknowledgments}

%\bibliographystyle{IEEEtran} %plain, alpha, amsplain, amsalpha (anvaa eshe)
%\bibliography{../../modeling_ref}


\end{document}